\documentclass[aps,superscriptaddress,nofootinbib,eqsecnum,prd,10pt,notitlepage,twocolumn]{revtex4}
\usepackage{amssymb}
\usepackage{amsmath}
\usepackage{natbib}
\usepackage{epsfig}
\usepackage{graphicx}
\usepackage{dcolumn}
\usepackage{bm}
\usepackage[english]{babel}
\usepackage[latin1]{inputenc}
\usepackage{eucal}
\usepackage{hyperref}
\usepackage{verbatim}
\usepackage{latexsym}
\usepackage{xcolor}
\usepackage{color}
\usepackage{soul}

\begin{document}

\title{NJL-type models in the presence of intense magnetic fields: the role of the regularization prescription}

\author{Sidney S. Avancini}
\email{sidney.avancini@ufsc.br}
\affiliation{Departamento de F\'{\i}sica, Universidade Federal de Santa
  Catarina, 88040-900 Florian\'{o}polis, Santa Catarina, Brazil }

\author{Ricardo L. S. Farias}
\email{ricardo.farias@ufsm.br}
\affiliation{Departamento de F\'{\i}sica, Universidade Federal de Santa Maria,
97105-900 Santa Maria, RS, Brazil}

\author{Norberto N. Scoccola}
\email{scoccola@tandar.cnea.gov.ar}
\affiliation{Department of Theoretical Physics, Comisi\'on Nacional de Energ\'ia At\'omica,
Av. Libertador 8250, 1429 Buenos Aires, Argentina}
\affiliation{Department of Theoretical Physics, Comisi\'on Nacional de Energ\'ia At\'omica,
\affiliation{Av. Libertador 8250, 1429 Buenos Aires, Argentina}
CONICET, Rivadavia 1917, 1033 Buenos Aires, Argentina}

\author{William R. Tavares}
\email{william.tavares@posgrad.ufsc.br}
\affiliation{Departamento de F\'{\i}sica, Universidade Federal de Santa
  Catarina, 88040-900 Florian\'{o}polis, Santa Catarina, Brazil }

\begin{abstract}
We study the regularization dependence of the Nambu-Jona--Lasinio model (NJL) predictions for some
properties of magnetized quark matter at zero temperature (and baryonic density) in the mean field
approximation.  The model parameter dependence for each regularization procedure is also analyzed
in detail. We calculate the average and difference of the quark condensates using different regularization
methods and compare with recent lattice results. In this context, the reliability of the different
regularization procedures is discussed.
\end{abstract}

\maketitle

\section{Introduction}
\label{sec1}
Many efforts have been dedicated to studying Quantum Chromodynamics (QCD) under extreme conditions
such as very high temperatures and densities~\cite{reviewsQCD}. One of the greatest challenges
at the present time is to understand
the physics that describe Quark Gluon Plasma (QGP), a new state of matter found experimentally,
that
corresponds to a thermalized color deconfined state of nuclear matter .  Many Heavy-Ion-Collisions
(HIC) experiments are under way, e.g., RHIC @ BNL, LHC @ CERN and others are
upcoming NICA @ JINR and FAIR @ GSI, to study this novel state of QCD matter and try to
obtain  some information that can help us to build a description of the unknown QCD
phase diagram. Although, from the theoretical point of view, a considerable amount of work
has been devoted to studying the phase
diagram of QCD, it still remains poorly understood. One of the main reasons is that
the energy range involved demands the calculation of QCD
in the non-perturbative regime, which is to date impracticable and
the ab initio lattice QCD approach has difficulties in dealing with the region
of moderately high densities due the ``sign problem''~\cite{karsch,nonaka}. In this
situation most of our
present knowledge about the QCD phase diagram arises from
the study of effective models, that offer the possibility of obtaining predictions
for regions that are no accessible through lattice techniques.

A topic that has attracted considerable attention in recent years is related to the fact
that in non-central heavy-ion collision strong magnetic fields can be generated.
In fact, they may reach strengths of the order of $10^{20}$ G~\cite{BinHIC1,BinHIC2}.
These strong magnetic fields, produced during the first instants after the collision,
can affect the QCD phases because they are of the order or higher than the QCD
scale $\Lambda_{QCD}^2$. More details of the recent advances in the understanding
of the phase structure and the phase transitions of hadronic matter
in strong magnetic fields can be found in recent reviews~\cite{andersenRMP,miranskyrev,ferrerepja}.
In recent years the number of articles dedicated to the study of the quark matter under
strong magnetic fields is immense and growing. The NJL model~\cite{reports} and its variations has a prominent role
in this context.
Since these models are non-renormalizable the calculation of observables within these
models demands always an appropriate regularization procedure to treat the divergent integrals.
The way that the divergencies are treated is of fundamental importance for
the results that are obtained in the calculations to be reliable.  In the literature several different
procedures have been used and many of them have serious problems which, in many situations,
ruin completely the conclusions of the calculations.
The scope of the present work is to
discuss an issue related to the application of the NJL model to the study of the properties
of the strongly interacting matter
in the presence of intense magnetic fields. We are particularly interested in the impact of
the use of different regularization procedures proposed in the literature within the SU(2)
version of the model. Namely, we discuss how the results for the behavior of the quark (u and d)
condensates as functions of the magnetic field depends on the way in which the NJL
is regularized. We pay special attention to the Magnetic Field Independent Regularization (MFIR)
proposed in Ref.~\cite{Ebert:1999ht,klimenko}. Such a scheme has been  recently  applied in several
works~\cite{Menezes:2008qt,sidneymfirsu3,mpi0plb,epja_runn,magsu2prc,mpi0PRD,prcIMC14} and, in particular,
it has been shown to avoid non-physical oscillations in the context of magnetized quark matter
in the presence of color superconductivity~\cite{scoccola_csc,prd16becbcs,mfirscs,GR_proceed}.
The procedure follows the steps of the dimensional regularization prescription of QCD,
performing a sum over all Landau levels in the vacuum term. In this procedure we can isolate the
divergence into a term that has the form of the zero magnetic field vacuum energy and that can be
regularized by different regularization schemes. As a criteria to determine which
regularization procedure is more appropriate to describe magnetized quark matter,
we confront our NJL results with recent
simulations of QCD on the lattice
(implemented at zero baryonic densities)~\cite{lattice,fodor,bali,Bali:2012zg}.
One of the main objectives of this work is to
clarify these
issues showing the appropriate way to be followed in order to obtain reliable results
in the calculations of physical quantities
using non-renormalizable models.

The paper has been organized as follows: in Sec.~\ref{sec2} we evaluate the quark condensates
within the NJL model in the presence of a constant magnetic field. In Sec.~\ref{sec3} we discuss
the MFIR regularization scheme, in Sec.~\ref{sec4} in the context of the MFIR
and non-MFIR (nMFIR) schemes the non-covariant and covariant regularization
schemes have been applied in the calculation of the quarks condensates using NJL model
in presence of strong magnetic fields and confronted with lattice results.
Finally, we conclude in Sec.~\ref{sec5}. We include two appendix (\ref{app1} and \ref{app2}) 
containing some details about
the magnetic field independent regularization procedure and the model parametrization
for each regularization scheme.


\section{Quark condensates within the NJL model in the presence of
a constant magnetic field}
\label{sec2}

Our starting point is the Euclidean effective action of the NJL model in the
presence of an external electromagnetic field. It reads:
\begin{eqnarray}
S_E &=& \int d^4 x
\left\{
\bar \psi (- i \gamma_\mu D_\mu + m_0) \psi
- G\left[(\bar \psi \psi)^2  \right.\right.\nonumber\\
&+&\left.\left.  (\bar \psi i\tau \gamma_{5}\psi)^2 \right]
\right\},
\end{eqnarray}
where the Euclidean $\gamma$ matrices satisfy
$\{ \gamma_\mu, \gamma_\nu\} = - 2 \delta_{\mu \nu}$~\cite{ripka},
$m_0$ is the current quark mass and $G$ is a coupling
constant. The coupling of the quarks to the electromagnetic field
${\cal A}_\mu$ is implemented  by the covariant derivative
$D_{\mu}=\partial_\mu - i q_f {\cal A}_{\mu}$ where $q_f$
represents the quark electric charge ($q_u/2 = -q_d = e/3$). We
consider a  static and constant magnetic field in the $3$-direction,
${\cal A}_\mu=\delta_{\mu 2}\ x_1\ B$. Since the model
under consideration is not renormalizable, a regularization scheme
needs to be specified. As it will be discussed below this introduces
an additional parameter $\Lambda$.
Together $m_0$, $G$ and $\Lambda$ form a set of three
parameters that completely determine the model. These parameters
are usually fixed in order to reproduce the empirical values in the
vacuum of the pion mass $m_\pi$, the pion decay constant $f_\pi$,
and the average quark condensate $\bar \Phi_0 = (<\bar u u >_0 + <\bar d d>_0)/2$.
Whereas the physical values
$m_\pi = 138.0$ MeV and $f_\pi = 92.4$ MeV, are known quite accurately,
the uncertainties for the quark condensate are rather large. Limits
extracted from sum rules are $190 \mbox{MeV} < -\bar \Phi_0^{1/3} < 260$
MeV at a renormalization scale of 1 GeV\cite{Dosch:1997wb}, while typically
lattice calculations yield
$\bar \Phi_0^{1/3} = - 231 \pm 8 \pm 6$ MeV \cite{Giusti:1998wy}
(see e.g. Ref.\cite{McNeile:2005pd} for some other
lattice results). In order to test the stability of our results
we will consider parametrizations leading to quark condensates in the
range $220$ MeV $< -\bar \Phi_0^{1/3} < 260$  MeV.

As it is well-known the presence of a constant magnetic field in the 3-direction
leads to a quantization of the momentum in the 1-2 plane. Thus, the
free energy in the mean field approximation can be obtained from
the one in the absence of magnetic field
\begin{equation}
{\cal F} = \frac{(M - m_0)^2}{4 G} - N_c \sum_{f,s}
\int \frac{d^4 p}{(2 \pi)^4}
\ln \left[ p^2 + M^2 \right] \label{free4} ,
\end{equation}
by using the replacement 
\begin{eqnarray}
\vec p\ ^2 &&\rightarrow \,\, p_3^2 + 2 k |q_f| B \nonumber\\
\sum_s \int \frac{d^4 p}{(2 \pi)^4} &&\rightarrow \,\,
\frac{|q_f|B}{2\pi}\int_{-\infty}^\infty \frac{dp_3}{2 \pi}
\int_{-\infty}^\infty \frac{dp_4}{2 \pi} \sum_{k=0}^\infty \alpha_k \label{presc}~,\nonumber \\
\end{eqnarray}
where $N_c=3$ is the number of colors, $f=u,d$ runs over the quark flavors and $s$
stands for the
spin label.
In addition, $M=m_0-2G \left\langle \bar{\psi} \psi \right\rangle$, is the
dressed quark mass, $k$ is the index associated with Landau levels (LL's) and
$\alpha_k = 2 -\delta_{k0}$ is the degeneracy factor.
The resulting expression is
\begin{eqnarray}
{\cal F} &=& \frac{(M - m_0)^2}{4 G} -N_c \sum_f
\frac{|q_f| B}{2\pi} \sum_{k=0}^\infty \alpha_k \ \nonumber \\&\times&
\int_{-\infty}^\infty \frac{dp_3}{2 \pi}
\int_{-\infty}^\infty \frac{dp_4}{2 \pi}
\ln \left[ p_4^2 + p_3^2 + 2 k|q_f| B + M^2 \right]\nonumber\\
\label{free}
\end{eqnarray}
from which the associated gap equation can be obtained from the condition
$\partial {\cal F}/\partial M = 0$. As expected
expression Eq.(\ref{free}) is  divergent and, thus, some regularization
scheme is required in order to proceed.
At this point we introduce another approach which is commonly used in the literature.
Instead of using directly Eq.(\ref{free4}) we firstly perform the integration in $p_4$
obtaining for the free energy:
\begin{equation}
{\cal F} = \frac{(M - m_0)^2}{4 G} - N_c \sum_{f,s}
\int \frac{d^3 p}{(2 \pi)^3}
\sqrt{ p^2 + M^2 } \label{free3} ~.
\end{equation}
Now, the replacement for obtaining the magnetized free energy, Eq.(\ref{presc}), is modified to:
\begin{equation}
\sum_s \int \frac{d^3 p}{(2 \pi)^3} \rightarrow \,\,
\frac{|q_f|B}{2\pi}\int_{-\infty}^\infty \frac{dp_3}{2 \pi} \sum_{k=0}^\infty \alpha_k ~.
\end{equation}
Therefore, one obtains:
\begin{equation}
{\cal F} = \frac{(M - m_0)^2}{4 G} - N_c \sum_f
\frac{|q_f| B}{2\pi} \sum_{k=0}^\infty  \alpha_k \  \int_{-\infty}^\infty \frac{dp_3}{2 \pi}
E_{p_3,k}~,
\label{f3d}
\end{equation}
where $E_{p_3,k} = \sqrt{p_3^2 + 2 k|q_f| B + M^2}$, this expressions is also
ultraviolet divergent  and
some regularization procedure has to be specified. The regularization procedure
will be discussed in detail
in the forthcoming sections.

As mentioned in the Introduction, our aim here is to compare the dependence of the quark
condensates  on the magnetic field with the existing lattice results, where the condensates are
calculated within the NJL model using different regularizations.
In particular, in Table 1 of Ref.\cite{Bali:2012zg} lattice data for the quantities
$\bar \Sigma = (\Sigma_u + \Sigma_d)/2$ and $\Sigma^- = \Sigma_u - \Sigma_d$,
where
\begin{eqnarray}
\Sigma_f(B,T) &=& \frac{2 m_0}{D^4} \left[ \Phi^f_{B,T} - \Phi^f_{0,0} \right] + 1
\ ,
\label{sigma}
\end{eqnarray}
are listed. Since here we are only interested in the case $T=0$, we will drop the second index
in what follows. In Eq.(\ref{sigma}), $\Phi^f \equiv <\bar f f>$ is the quark condensate associated to the flavor $f$
and the constant $D$, taken to be $D =(135 \times 86)^{1/2}$ MeV as in Ref.\cite{Bali:2012zg}, is introduced just
for dimensional reasons. In addition, as in the latter reference, we are working in the isospin
limit $m_u = m_d = m_0$.
We are mainly interested in the behavior of the change of the condensate due to the magnetic field.
Thus, following Ref.\cite{Bali:2012zg} we define
\begin{equation}
\Delta \Sigma_f(B) = \Sigma_f(B) -  \Sigma_f(0)
\end{equation}
in terms of which the change in the average of the flavor condensates is
\begin{eqnarray}
\Delta \bar \Sigma \equiv \frac{\Delta\Sigma_u(B) + \Delta\Sigma_d(B)}{2} = - \frac{2 m_0}{D^4}
\left( \bar \Phi_B -  \bar \Phi_0 \right)  ~,
\end{eqnarray}
where  $\bar \Phi_B = ( \Phi^u_B + \Phi^d_B)/2 $ indicates the average quark
condensate for arbitrary magnetic field $B$.
In order to compare with lattice results we calculate this quantity using
the condensates as
evaluated in the NJL. Since in this model the average quark condensate is
related to the dressed
quark mass as
\begin{eqnarray}
\bar \Phi = - \frac{M-m_0}{4 G} ~,
\label{cond}
\end{eqnarray}
we get
\begin{eqnarray}
 \Delta \bar \Sigma  = \frac{m_0}{D^4} \frac{M_B -  M_0}{2 G} ~,
\label{condmass}
\end{eqnarray}
where $M_0$ if the constituent quark mass evaluated in the absence of magnetic field.

Using the definition Eq.(\ref{sigma}), we introduce the difference between the condensates
\begin{eqnarray}
\Sigma^{-} = \Sigma_u - \Sigma_d = \frac{2m_0}{D^4}\left(\Phi_{B}^u-\Phi_{B}^d\right)\label{dif} ~.
\end{eqnarray}
We recall here that the definition of $\Phi_{B}^f$ depends on the regularization procedure adopted as 
will be explained in the following sections.

The parameters used in our calculations for the different
regularization schemes to be discussed in detail in the following
sections are given in Table~\ref{resfitb}. They were determined
by fitting the pion mass and its decay constant to their empirical
values $m_\pi = 138$ MeV and $f_\pi = 92.4$ MeV, respectively, and
the average quark condensate $\bar \Phi_0$ to values within the
phenomenological range $-\Phi_0^{1/3} = 220 - 260$ MeV.
\begin{widetext}
\begin{center}
\begin{table}[h]
\begin{tabular}{cccccc}
\hline
\hspace*{0.5cm} Regulation type \hspace*{0.5cm}     &\hspace*{0.5cm} $-\bar  \Phi_0^{1/3}$\hspace*{0.5cm} & \hspace*{0.5cm} $M_0$   \hspace*{0.5cm}
& \hspace*{0.5cm} $G \Lambda^2$\hspace*{0.5cm}  & \hspace*{0.5cm}$\Lambda$ \hspace*{0.5cm}   & \hspace*{0.5cm}$m_0$\hspace*{0.5cm} \\
 \cline{2-6}
                      & MeV          &  MeV       &                 & MeV          & MeV \\
\hline
   Lorenztian N=5            &  245.0       &  428.85   & 2.333   & 569.52   &  5.455  \\
                      &  260.0       & 286.19    & 1.860   & 681.38   &  4.552 \\
\hline
   Woods-Saxon $\alpha =0.1$   &  245.0       &  399.48   & 2.316   & 588.07   &  5.452  \\
                      &  260.0       & 285.44    & 1.923   & 693.77    &  4.552  \\
\hline
   Gaussian                         &  250.0       &  394.52   & 2.236   & 598.53   &  4.456  \\
                      &  260.0       & 311.47    & 1.994   & 675.26    &  3.956  \\
\hline
   Fermi-Dirac $\alpha =0.01\Lambda$   &  245.0       &  333.53   & 2.188   & 626.34   &  5.438  \\
                      &  260.0       & 270.18    & 1.954   & 719.17    &  4.548  \\
\hline
  3D cutoff           & 241.0        & 390.32    & 2.404   & 591.6   &  5.723  \\
                      & 260.0        & 270.14    & 1.954   & 719.23   &  4.548  \\
\hline
   Proper Time        & 220.0        & 224.17    & 4.001   &  886.62  &  7.383   \\
                      & 260.0        & 191.70    & 3.608   & 1164.10   &  4.516   \\
\hline
4D cutoff             & 220.0        & 305.58    & 4.568   &  807.83   &  7.449  \\
                      &  260.0       &  222.72   &  3.719  & 1094.76   &  4.531  \\
                      \hline
Pauli Villars             & 220.0        & 313.20    & 3.337   &  681.84   &  7.453  \\
                      &  260.0       &  224.67   &  2.688  & 926.57   &  4.532  \\

\hline
\end{tabular}
\caption{ Parametrizations of the NJL model for the different regularization schemes.}
\label{resfitb}
\end{table}
\end{center}
\end{widetext}
\section{Magnetic Field Independent regularization - MFIR}
\label{sec3}
The magnetic field independent regularization (MFIR) was developed in Ref.\cite{Ebert:1999ht} and
there it was shown that it is possible to separate a divergent vacuum contribution
from a finite magnetic field contribution. This was achieved in Ref.\cite{Menezes:2008qt}
by using the dimensional regularization method.
In this section we will study this regularization method both in the case where all components
of the quark four-momentum are treated on an equal footing and after an integration in $p_4$. For this purpose
it is convenient to start
from the derivative with respect to dressed mass of the (unregularized) free energy Eq.(\ref{free}).
Namely,
\begin{eqnarray}
\frac{\partial{{\cal F}}}{\partial M} &=&
\frac{M - m_0}{2G} - 2 M N_c ~  \tilde{I}  ~,~\nonumber \\
\tilde{I} & = & \sum_f \frac{|q_f|B}{2 \pi} \sum_{k=0}^\infty \alpha_k
\int_{-\infty}^\infty \frac{dp_4}{2 \pi}\nonumber\\
&\times&
 \int_{-\infty}^\infty \frac{dp_3}{2 \pi} \frac{1}{p_4^2 + p_3^2 + 2 k |q_f| B + M^2} ~.
\label{gap4DD}
\end{eqnarray}
At this stage we add and subtract the contribution in the absence of magnetic field.
We get then,
\begin{equation}
\tilde{I} = \left[ I_1 + \sum_f I_f \right] ~,
\label{general4D}
\end{equation}
where
\begin{eqnarray}
I_1  = 4 \int \frac{d^4p}{(2\pi)^4} \frac{1}{p^2 + M^2}\label{I1vac}
\end{eqnarray}
and
\begin{eqnarray}
I_f &=&  \int_{-\infty}^\infty \frac{dp_4}{2 \pi}
\int_{-\infty}^\infty \frac{dp_3}{2 \pi} \left[\frac{|q_f|B}{2 \pi}
\sum_{k=0} \alpha_k \right.\nonumber \\
&\times &\left. \frac{1}{p_3^2 + p_4^2 + 2 k |q_f| B + M^2}
-   2 \int_{-\infty}^\infty \frac{dp_1}{2 \pi} \right.\nonumber \\
&\times& \left.\int_{-\infty}^\infty \frac{dp_2}{2 \pi}
\frac{1}{p_1^2 + p_2^2 + p_3^2 + p_4^2 + M^2}\right] ~.
\label{if}
\end{eqnarray}
Interestingly, while $\tilde{I}$  in Eq.~\ref{gap4DD} is divergent and requires some type of
regularization procedure,  $I_f$ in Eq.~\ref{if} is finite. In fact, as shown in the Appendix-\ref{app1},
one has
\begin{equation}
I_f = \frac{M^2}{8\pi^2} \ \eta(x_f) ~,
\label{iffinal}
\end{equation}
where  $\eta(x)$ is give by:
\begin{eqnarray}
\eta(x) = \frac{\ln \Gamma(x)}{x} - \frac{\ln{2\pi}}{2x} + 1 - \left(1-\frac{1}{2x}\right) \ln x ~,
\label{eta}
\end{eqnarray}
where $x_f = M^2/(2 |q_f| B)$.
Therefore, Eq.(\ref{gap4DD}) can be casted into the form
\begin{equation}
\frac{\partial{{\cal F}}}{\partial M} = \frac{M - m_0}{2 G}  - 2 M N_c I_1 -
\frac{Nc}{4\pi^2} M^3 \sum_f  \eta(x_f) ~,
\label{general4D-2}
\end{equation}
from which the explicit form of the regularized free energy can be obtained by integration. However,
from the way it has been derived here, we see that
any covariant regularization method can be used to treat the vacuum term as well.
For example, in
Ref.\cite{Ebert:1999ht}  a 4D sharp cutoff was used.

For the alternative form of the free energy, Eq.(\ref{f3d}), proceeding analogously
as above, one obtains:

\begin{equation}
\frac{\partial{{\cal F}}}{\partial M} =
\frac{M - m_0}{2G} - 2 M N_c~\tilde{I}^{3D} ~,
\label{general3D}
\end{equation}
\begin{equation}
\tilde{I}^{3D} = \sum_f  \frac{|q_f|B}{4 \pi}
\sum_{k=0} \alpha_k \int_{-\infty}^\infty \frac{dp_3}{2 \pi}
 \frac{1}{\sqrt{p_3^2 + 2 k |q_f| B + M^2}}       ~, \label{gap3DD}
\end{equation}
where, after adding and subtracting the non-magnetic vacuum term one obtains:
\begin{equation}
 \tilde{I}^{3D} = \left[ I_1^{3D} + \sum_f I_f^{3D} \right] ~,
\end{equation}
where
\begin{eqnarray}
I_1^{3D}  = 2 \int \frac{d^3p}{(2\pi)^3} \frac{1}{\sqrt{p^2 + M^2}}\label{I3vac}
\end{eqnarray}
and
\begin{eqnarray}
I_f^{3D} &=&  \int_{-\infty}^\infty \frac{dp_3}{2 \pi} \left[\frac{|q_f|B}{4 \pi}
\sum_{k=0} \alpha_k
 \frac{1}{\sqrt{p_3^2 + 2 k |q_f| B + M^2}}  \right.\nonumber \\
&& - \left.   \int_{-\infty}^\infty \frac{dp_1}{2 \pi}
\int_{-\infty}^\infty \frac{dp_2}{2 \pi}
\frac{1}{\sqrt{p_1^2 + p_2^2 + p_3^2  + M^2}}\right]
\label{if3D}  ~.\nonumber \\
\end{eqnarray}
The finite magnetic contribution, $I_f^{3D}$,  was obtained in\cite{Menezes:2008qt}, and coincides
with the expression given in Eq.(\ref{iffinal}):

The resulting derivative of the free energy, $\frac{\partial {\cal F} }{\partial M} $, can be
written as:
\begin{eqnarray}
\frac{\partial{{\cal F}}}{\partial M} = \frac{M - m_0}{2 G} - 2 M N_c  I_1^{3D} -  \frac{N_c}{{4}\pi^2}
M^3
\sum_f \eta(x_f)  \label{gapeq3d} ~,\nonumber\\
\end{eqnarray}
where $I_1^{3D}$ is given in Eq.(\ref{i13d}) of the Appendix \ref{app2}.

\section{Regularization Procedures}
\label{sec4}

From the discussion in the previous section, it is clear that we have to specify a regularization
procedure in order to perform the calculation of any quantity within the NJL model. In fact,
this choice has to be considered a part of the model. In principle, we have two possibilities for
the regularization scheme to be used in the calculation of the condensate:\\
a) nMFIR regularization: in this case the gap equation, or equivalently,  the condensate
through Eq.(\ref{cond}) is calculated regularizing directly the expressions
$\tilde{I}$ or $\tilde{I}^{3D}$ given in Eq.(\ref{gap4DD}) and Eq.(\ref{gap3DD}). In this procedure
the magnetic and non-magnetic vacuum contributions are entangled and the consequences of this
choice will be addressed in this section.\\
b) MFIR regularization: in this procedure the gap equation is calculated regularizing only
the non-magnetic vacuum integrals $I_1$ or $I_1^{3D}$ in Eq.(\ref{general4D-2}) and Eq.(\ref{gapeq3d}). This
procedure separates exactly the finite magnetic term from the divergent non-magnetic one. \
Next, we will discuss the advantages and disadvantages of each regularization scheme. Here, we
emphasize that both procedures are largely utilized in the literature.
\subsection{Non-covariant regularizations}
We start with the case where the integral over $p_4$ was performed in the expressions of interest.
\subsubsection{Form factor regularizations}
 Firstly, we discuss how form factor regularizations are introduced within the nMFIR scheme.
In this kind of  regularization a form factor $U_\Lambda$ is introduced such that in Eq.(\ref{gap3DD})
\begin{eqnarray}
\sum_{k=0}^\infty  \int_{-\infty}^\infty \frac{dp_3}{2 \pi}  \rightarrow
\sum_{k=0}^\infty  \int_{-\infty}^\infty \frac{dp_3}{2 \pi} \ U_\Lambda(p_3^2 + 2 k|q_f| B)
\label{repff}
\end{eqnarray}
This particular procedure was used in the comparison of (P)NJL
results to lattice results indicated in Fig.3 of
Ref.\cite{Bali:2012zg}, and which leads to the statement that a
good agreement is only obtained for magnetic fields smaller that
about $0.3$ GeV$^2$. In addition, it is interesting to note that
non-physical oscillations might arise with this regularization
scheme, and these oscillations  are more evident in studies which
include color pairing interactions based on this kind of regularization. Interesting applications
of MFIR in this context
can be found in
 Refs.~\cite{scoccola_csc,prd16becbcs,mfirscs,GR_proceed}.
From the application of the replacement Eq.(\ref{repff})
in Eq.(\ref{gap3DD}), the gap equation $\frac{\partial \cal F} {\partial M}=0 $,
can be casted into the form

\begin{eqnarray}
M &=& m_0 + \frac{N_c}{2 \pi^2} \ G \ M
\sum_f |q_f| B \sum_{k=0}^\infty \alpha_k \nonumber \\
&\times&
\int_{-\infty}^\infty dp_3 \frac{U_\Lambda(p_3^2 + 2 k|q_f| B)}{E_{p_3,k}}
\label{gapff}
\end{eqnarray}

To solve this equation the specific form of $U_\Lambda$ has to be specified.
In principle, one might be tempted to use a simple step function
$\theta(x - \Lambda^2)$. However, this introduces strong unphysical
oscillations in the behavior of different quantities as functions of the magnetic
field.  To avoid these difficulties different smooth form factors
have been used in the literature. For example, in Refs.\cite{Gatto:2010pt,Frasca:2011zn} the
Lorenztian function
\begin{equation}
U_\Lambda^{(LorN)}(x) = \left[ 1 + \left(\frac{x}{\Lambda^2}\right)^{N}\right]^{-1}
\end{equation}
has been used.
Alternatively, in Ref.\cite{Fayazbakhsh:2010gc} Woods-Saxon (WS) type form factors
\begin{equation}
U_\Lambda^{(WS\alpha)}(x) = \left[ 1 + \exp\left(\frac{x/\Lambda-1}{\alpha}\right)\right]^{-1}
\end{equation}
have been used. It should be noted that all these form factors
include an additional parameter that controls their smoothness. To
choose the values of such parameter one has to take into account
that a too steep function gives rise to the unphysical
oscillations mentioned above and that a too smooth function leads
to values of the average quark condensate $\Phi_0$ which are quite
above the phenomenological range. Thus, the value $N=5$ is usually
chosen in the case of the Lorenztian form factor (Lor5)
while $\alpha = 0.1$ is taken for the case of Woods-Saxon
one (WS).

In Ref.\cite{noronha,ferrer} the authors introduce a Gaussian regulator (GR)
with momentum cutoff $\Lambda = 1$ GeV.
\begin{equation}
U_\Lambda^{(GR)}(x) = \exp \left(-\frac{x^2}{\Lambda^2}\right)
\end{equation}

In Ref.\cite{jaikumar2017} the authors use the following Fermi-Dirac-type smooth cutoff function

\begin{equation}
U^{FD}_\Lambda(x) = \frac{1}{2}\left[1-\tanh\left(\frac{\frac{x}{\Lambda}-1}{\alpha}\right)\right]
\end{equation}

where $\alpha=0.01$.

To evaluate the difference between the condensates in this scheme
we use the definition of the condensates $\Phi_{B}^f$ using form
factors:

\begin{eqnarray}
 \Phi_{B}^f=-2N_cM\frac{|q_f|B}{4\pi}\sum_{k=0}^{\infty}\alpha_k\int_{-\infty}^{\infty}\frac{dp_3}{2\pi}
 \frac{U_{\Lambda}(p_3^2+2k|q_f|B)}{E_{p_3,k}}\nonumber \\
\end{eqnarray}
\begin{figure*}[!ht]
\begin{center}
\includegraphics[width=0.45\linewidth,angle=0]{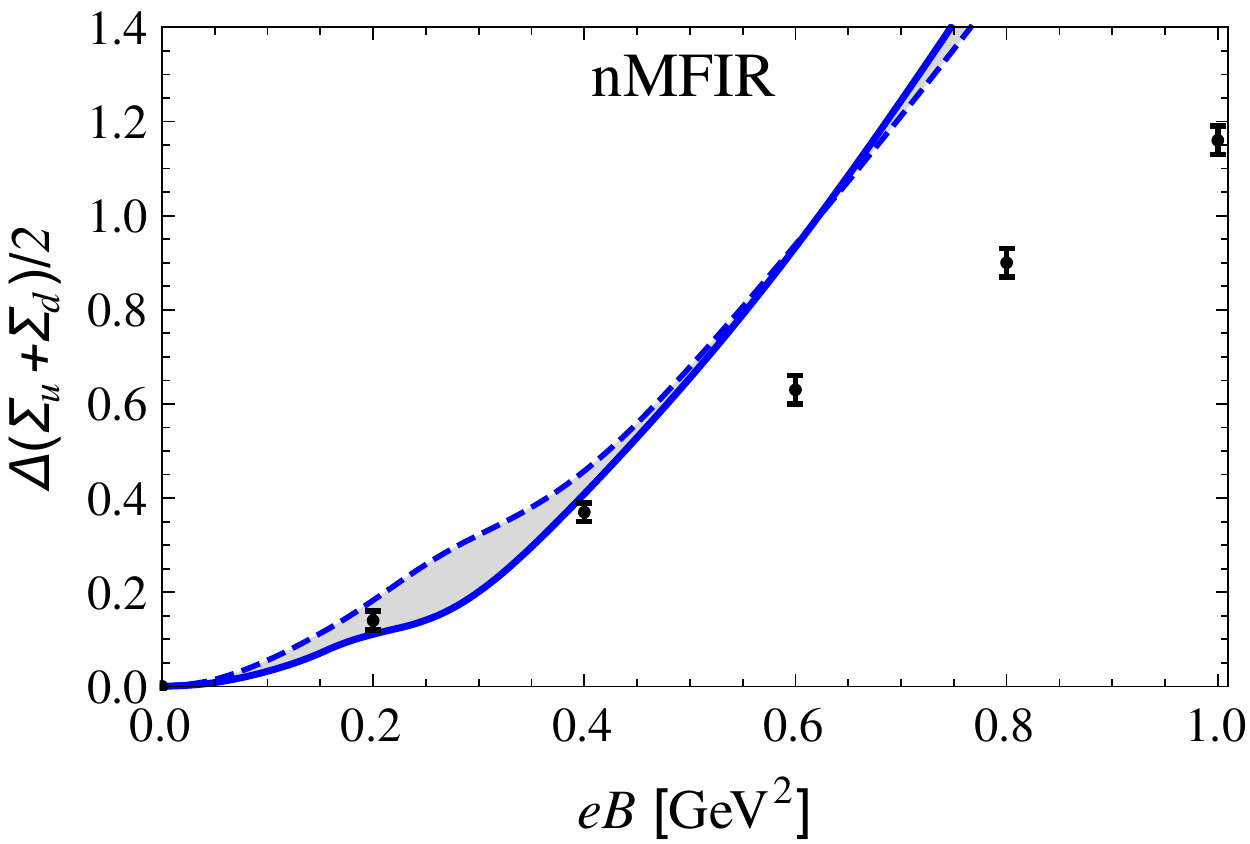}
\includegraphics[width=0.45\linewidth,angle=0]{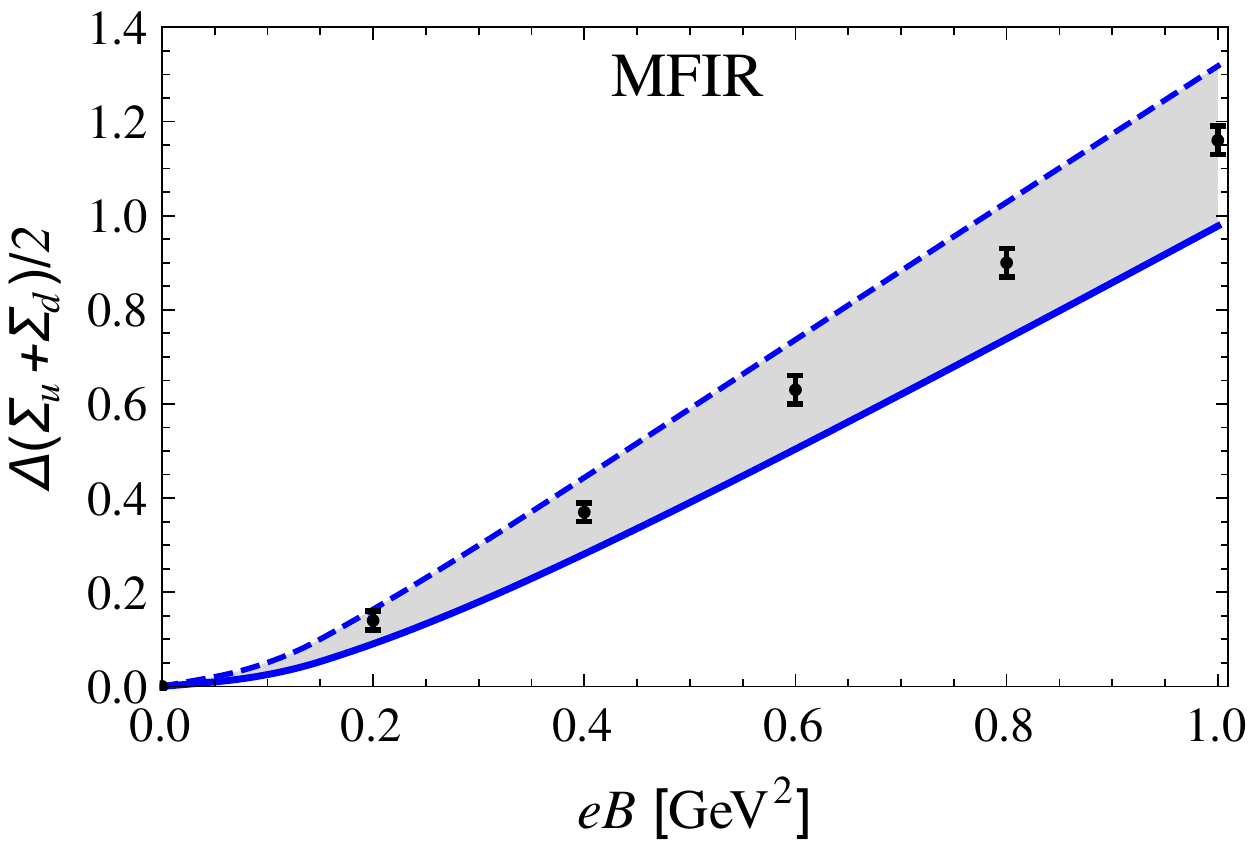}
\includegraphics[width=0.45\linewidth,angle=0]{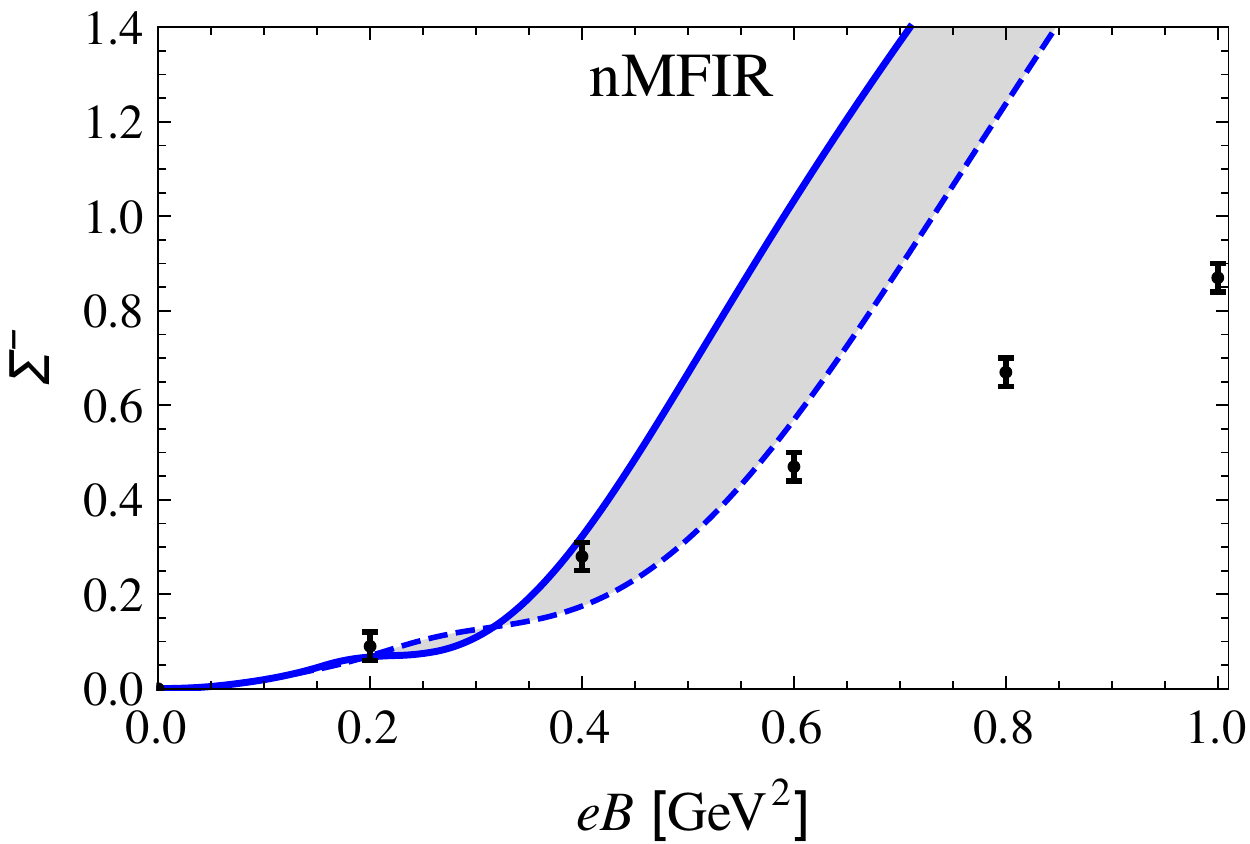}
\includegraphics[width=0.45\linewidth,angle=0]{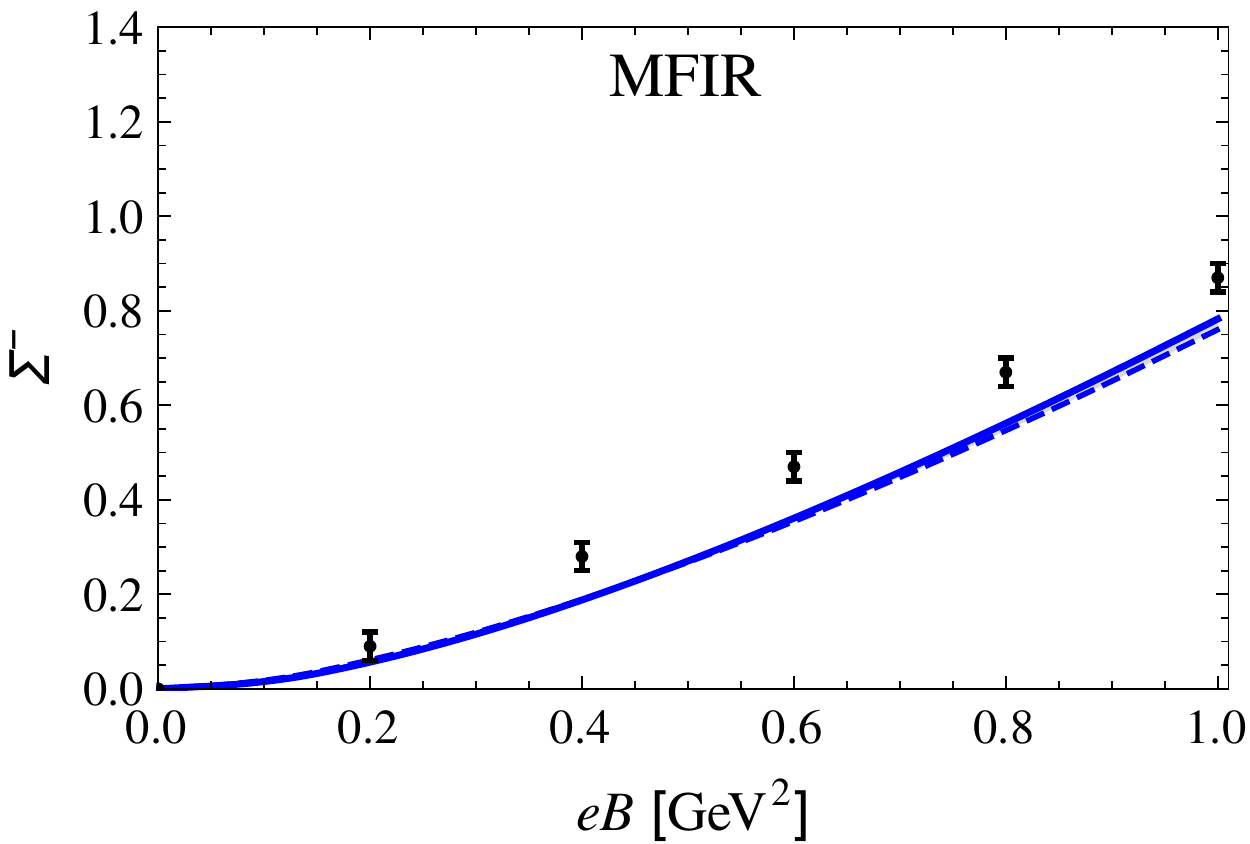}
\caption{{Results for the Lor5 form factor compared with
lattice results. Upper panels: Average flavor condensate as a
function of $eB$: nMFIR (left panel)  and  MFIR (right
panel). Lower panels: difference of the up and the down quark
condensates as a function of $eB$: nMFIR (left panel)  and
MFIR (right panel)} }
\label{fig1}
\end{center}
\end{figure*}

\begin{figure*}[!t]
\begin{center}
\includegraphics[width=0.45 \linewidth,angle=0]{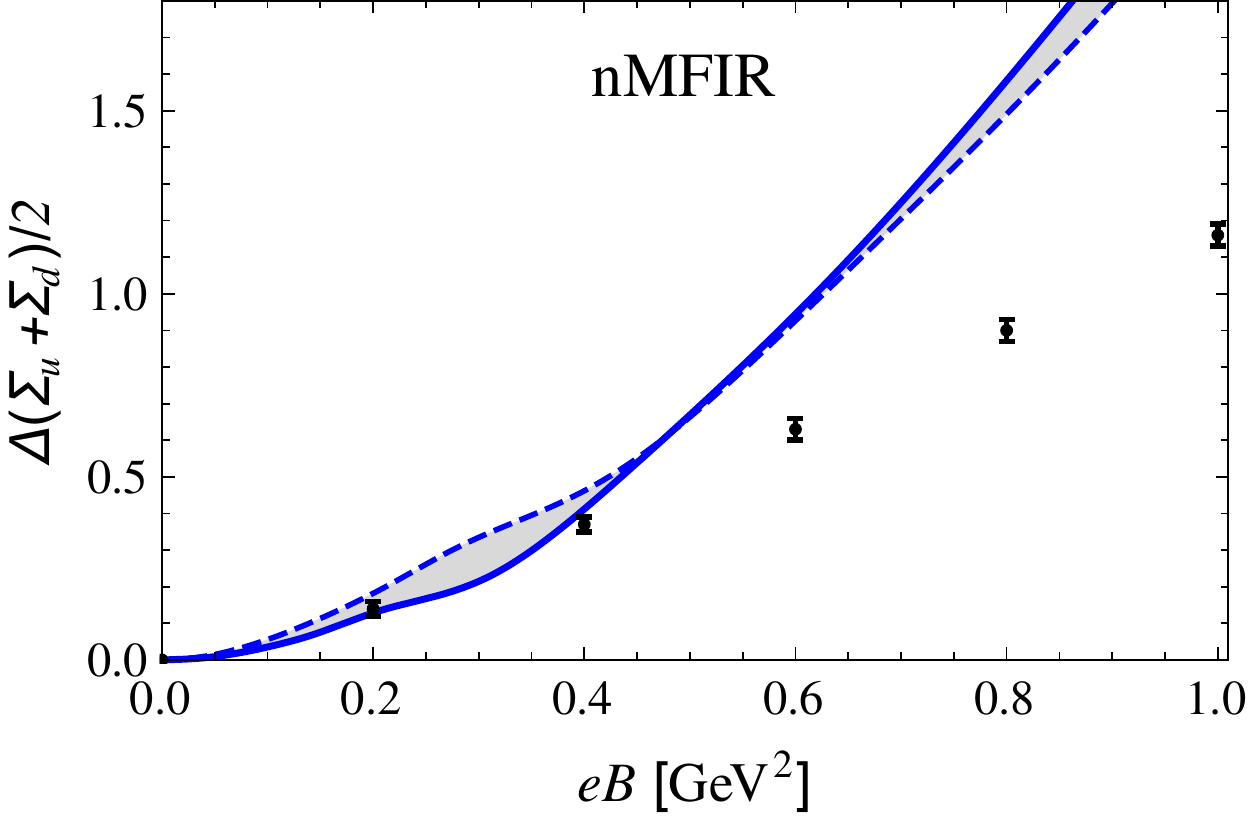}
\includegraphics[width=0.45 \linewidth,angle=0]{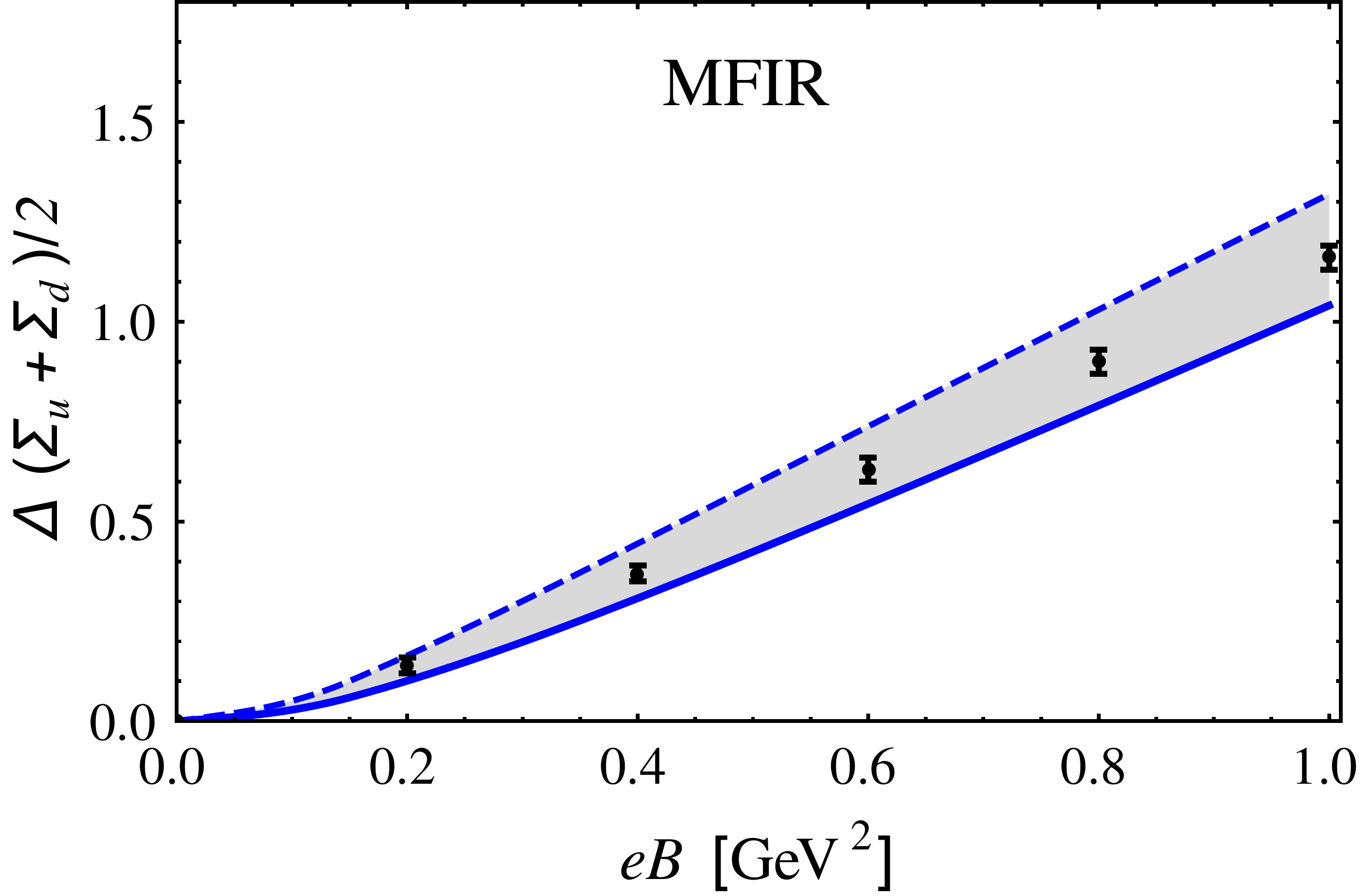}
\includegraphics[width=0.45 \linewidth,angle=0]{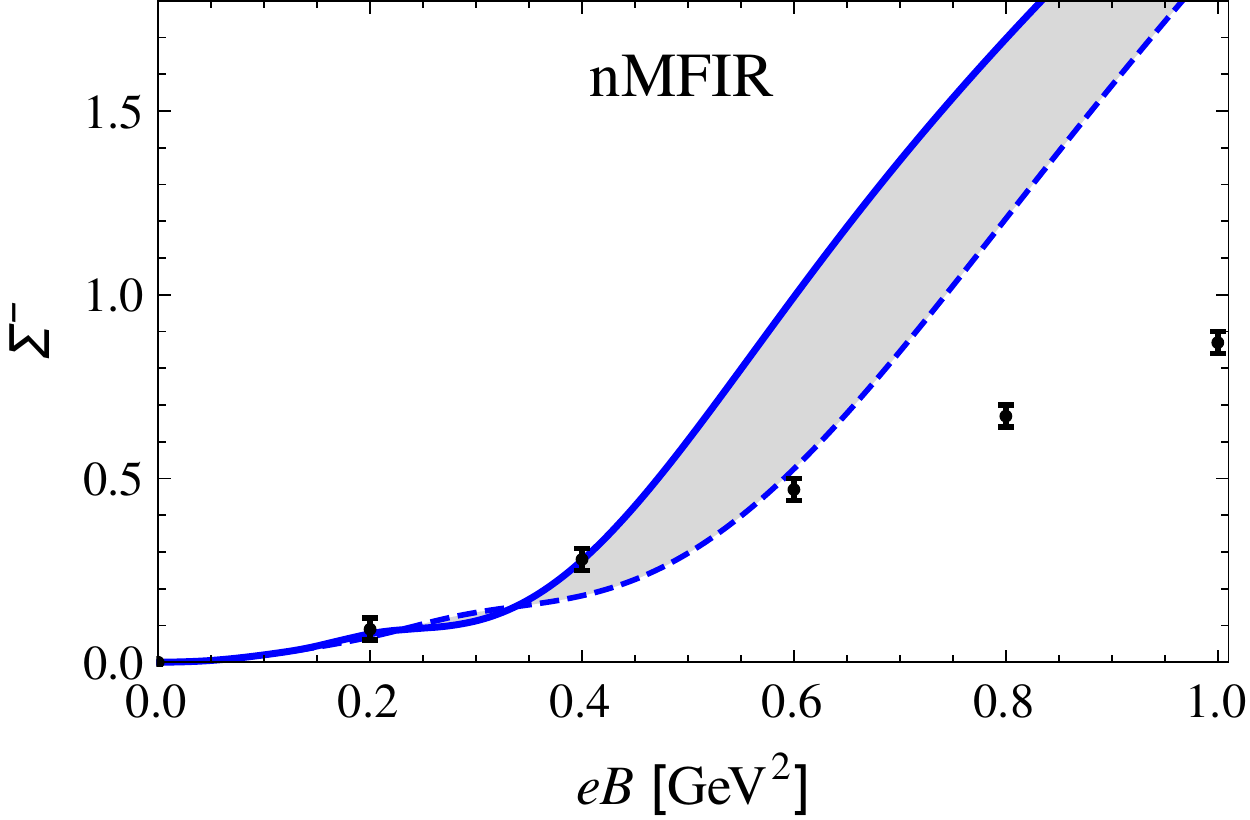}
\includegraphics[width=0.45 \linewidth,angle=0]{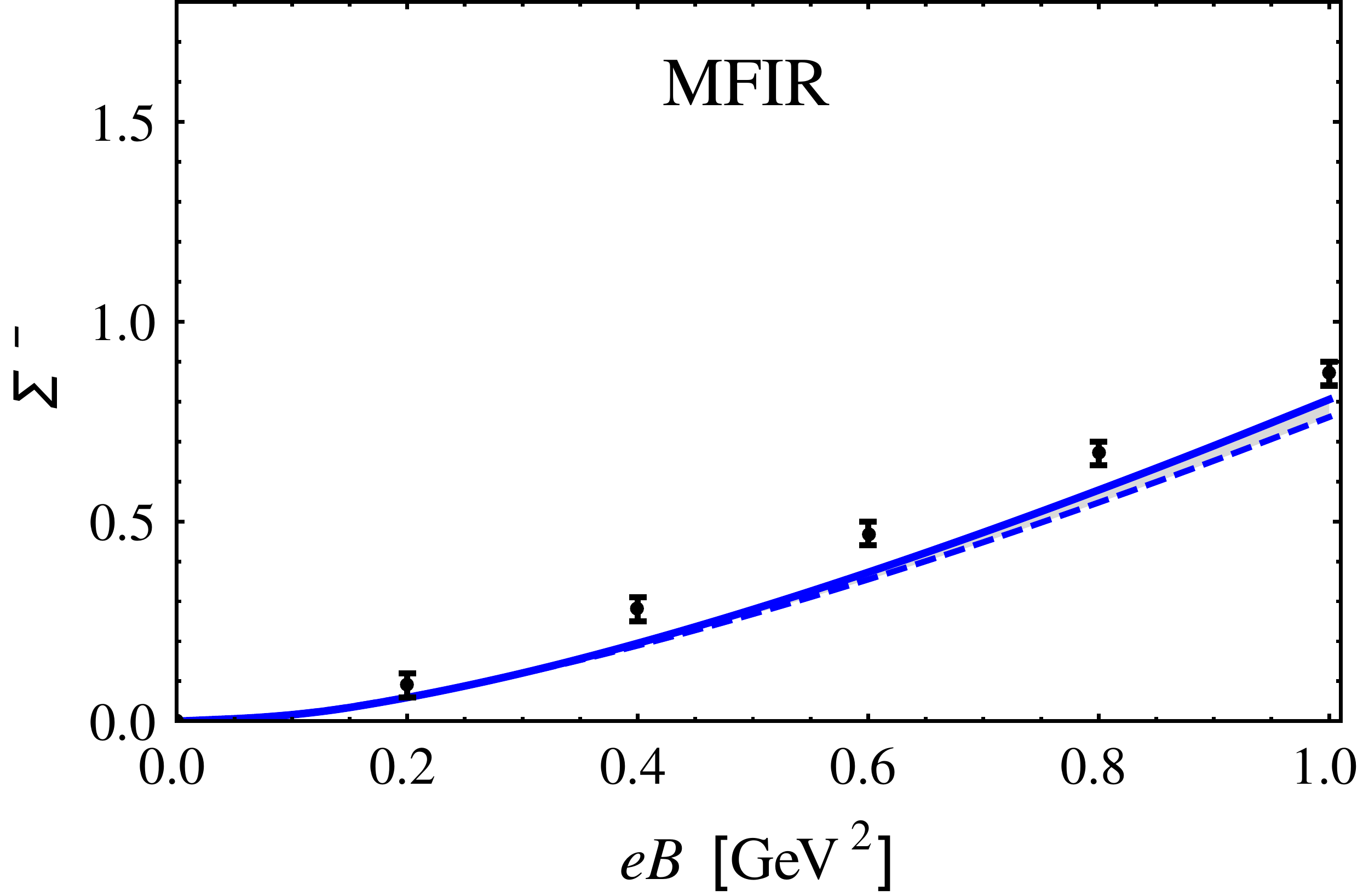}
\caption{{Results for the WS form factor with $\alpha = 0.1$
compared with lattice results. Upper panels: Average flavor
condensate as a function of $eB$: nMFIR (left panel)  and
MFIR (right panel). Lower panels: difference of the up and the
down quark condensates as a function of $eB$: nMFIR (left
panel) and MFIR (right panel)}}
\label{fig2}
\end{center}
 \end{figure*}

 \begin{figure*}[!t]
 \begin{center}
\includegraphics[width=0.45 \linewidth,angle=0]{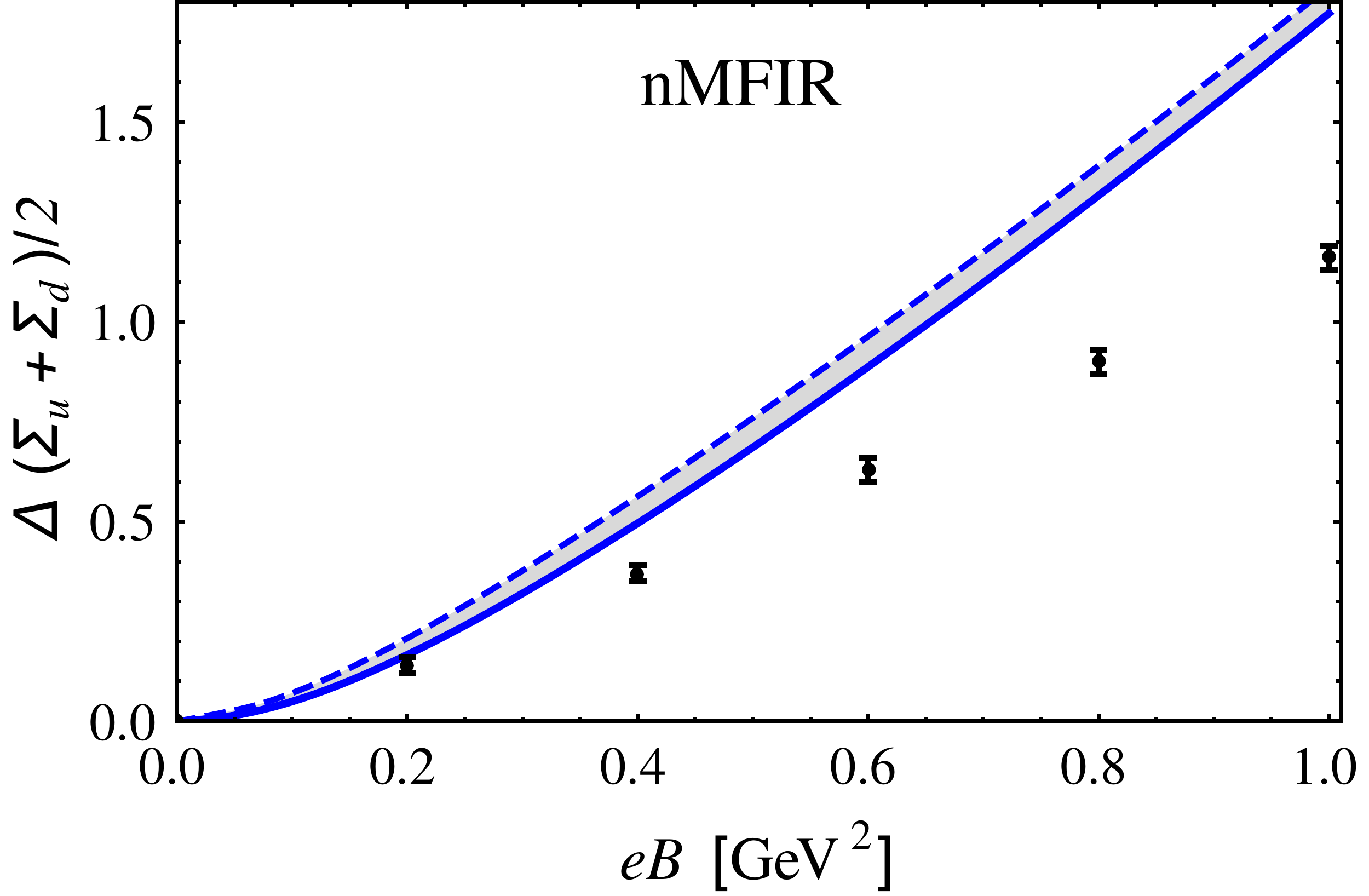}
\includegraphics[width=0.45 \linewidth,angle=0]{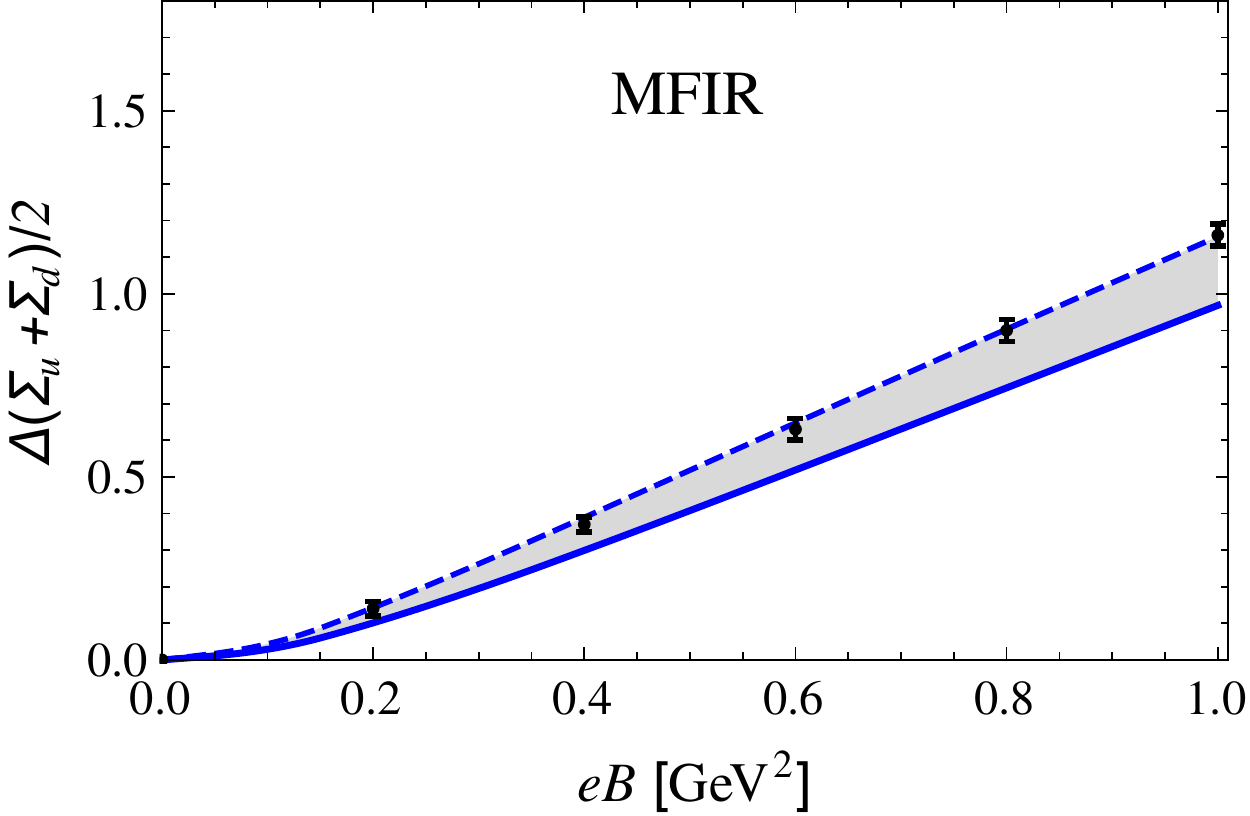}
\includegraphics[width=0.45 \linewidth,angle=0]{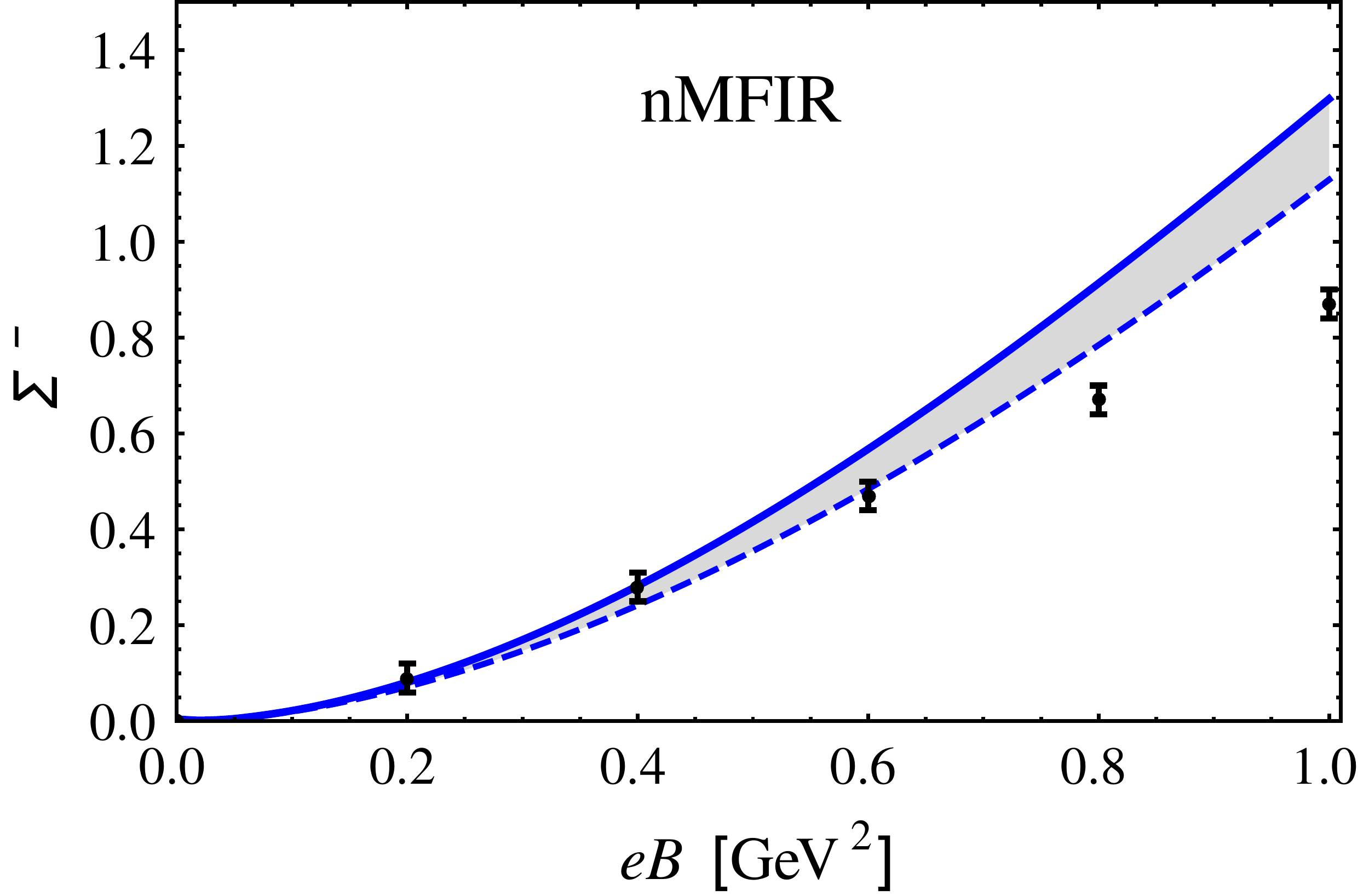}
\includegraphics[width=0.45 \linewidth,angle=0]{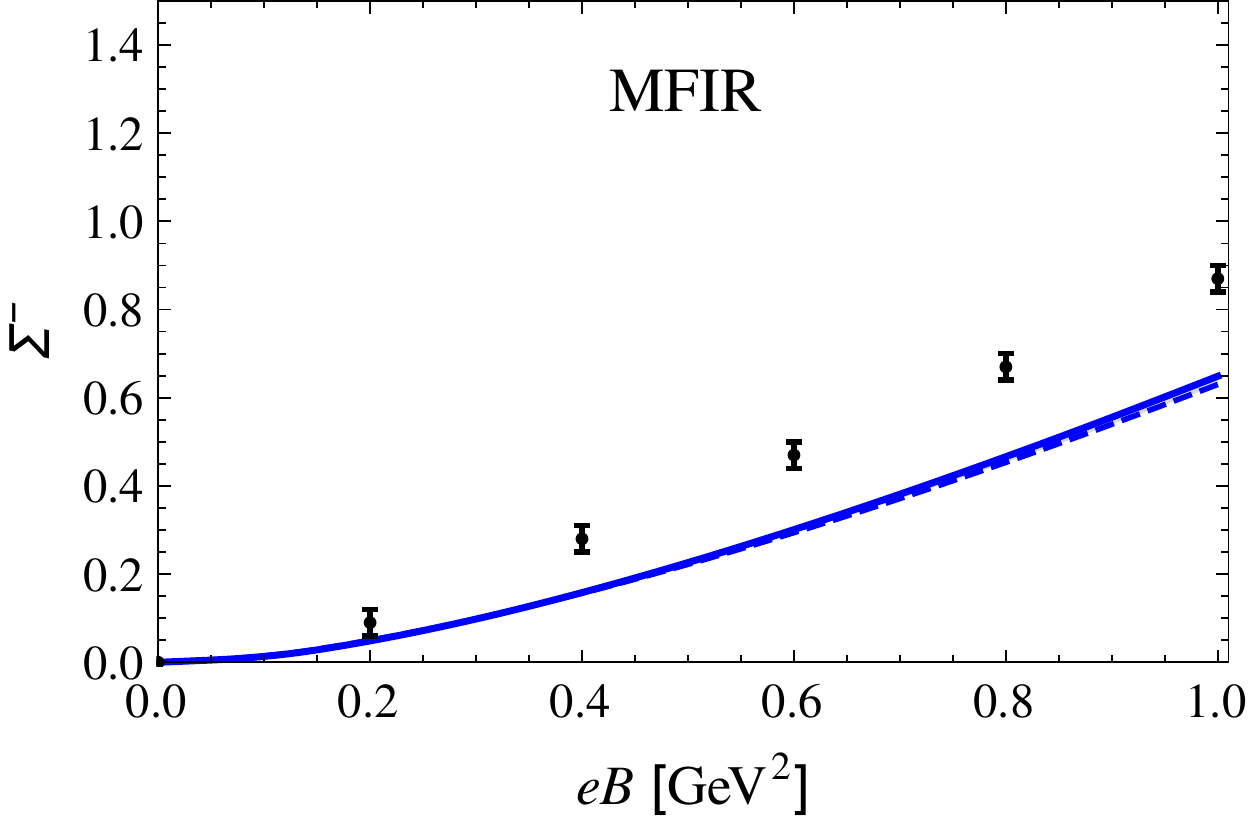}
\caption{{Results for the GR form factor compared with
lattice results. Upper panels: Average flavor condensate as a
function of $eB$: nMFIR (left panel) and MFIR (right
panel). Lower panels: difference of the up and the down quark
condensates as a function of $eB$: nMFIR (left panel)  and
MFIR (right panel)}}
\label{fig3}
\end{center}
 \end{figure*}

\begin{figure*}[!t]
\begin{center}
\includegraphics[width=0.45 \linewidth,angle=0]{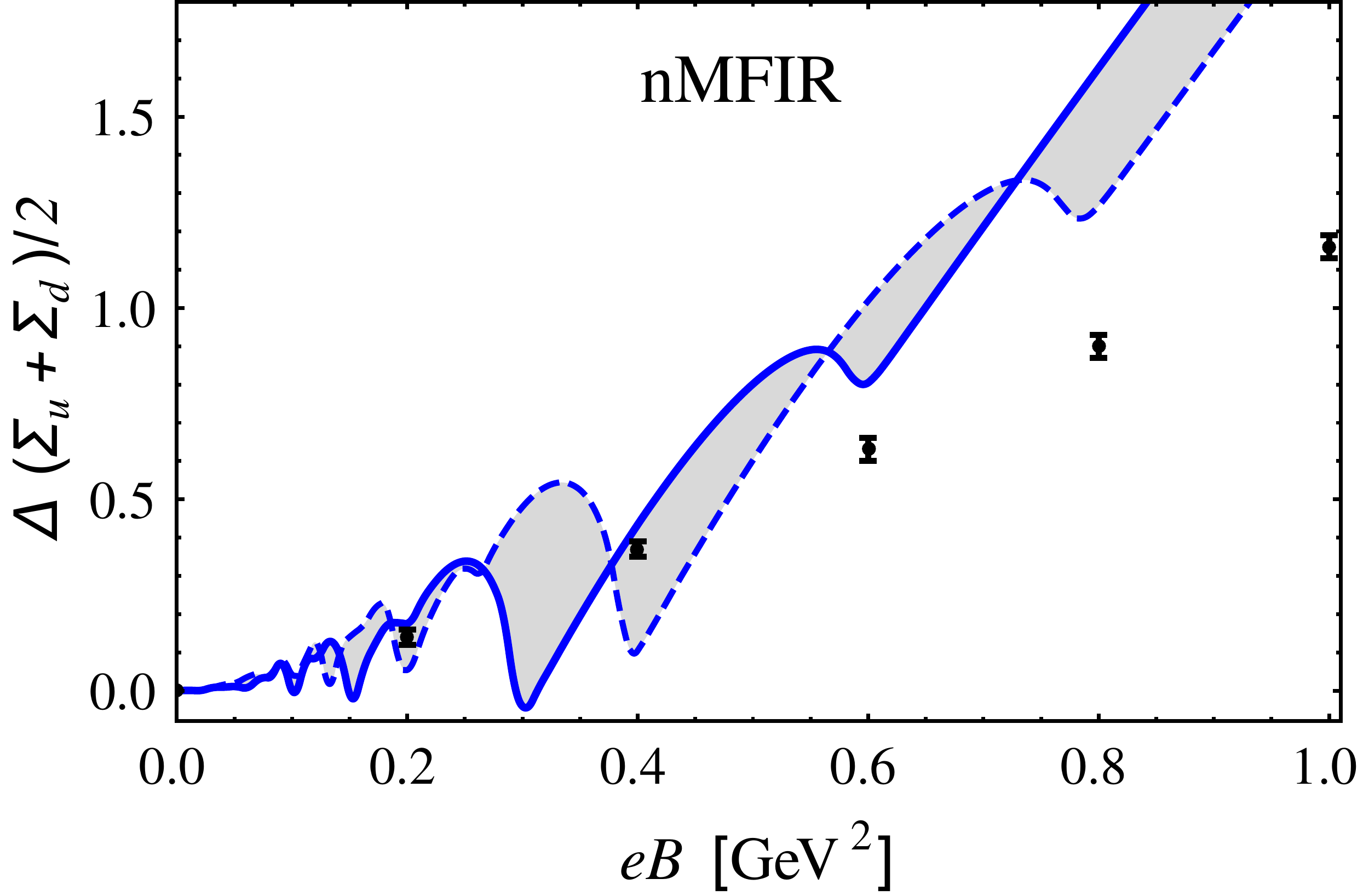}
\includegraphics[width=0.45 \linewidth,angle=0]{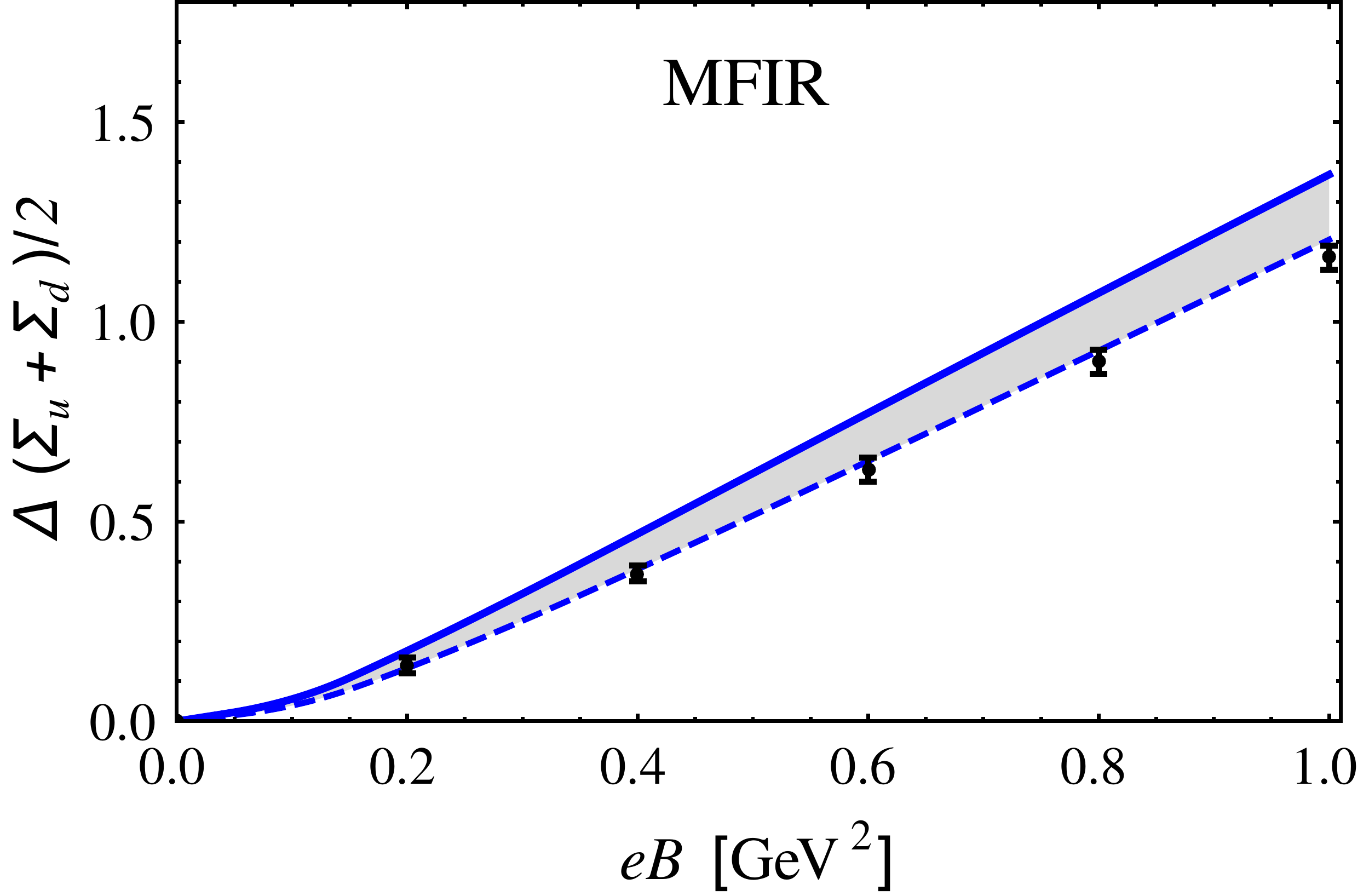}
\includegraphics[width=0.45 \linewidth,angle=0]{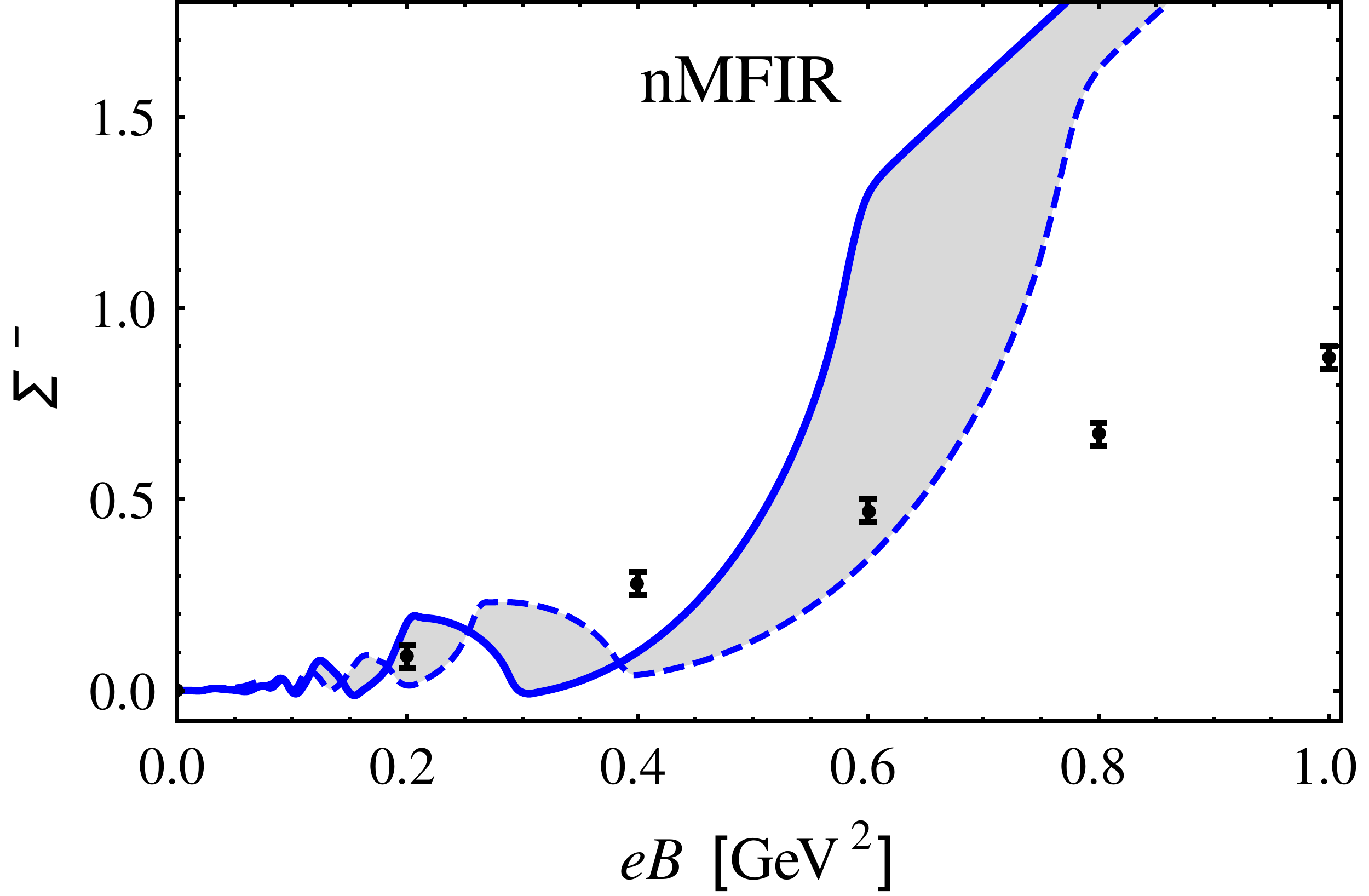}
\includegraphics[width=0.45 \linewidth,angle=0]{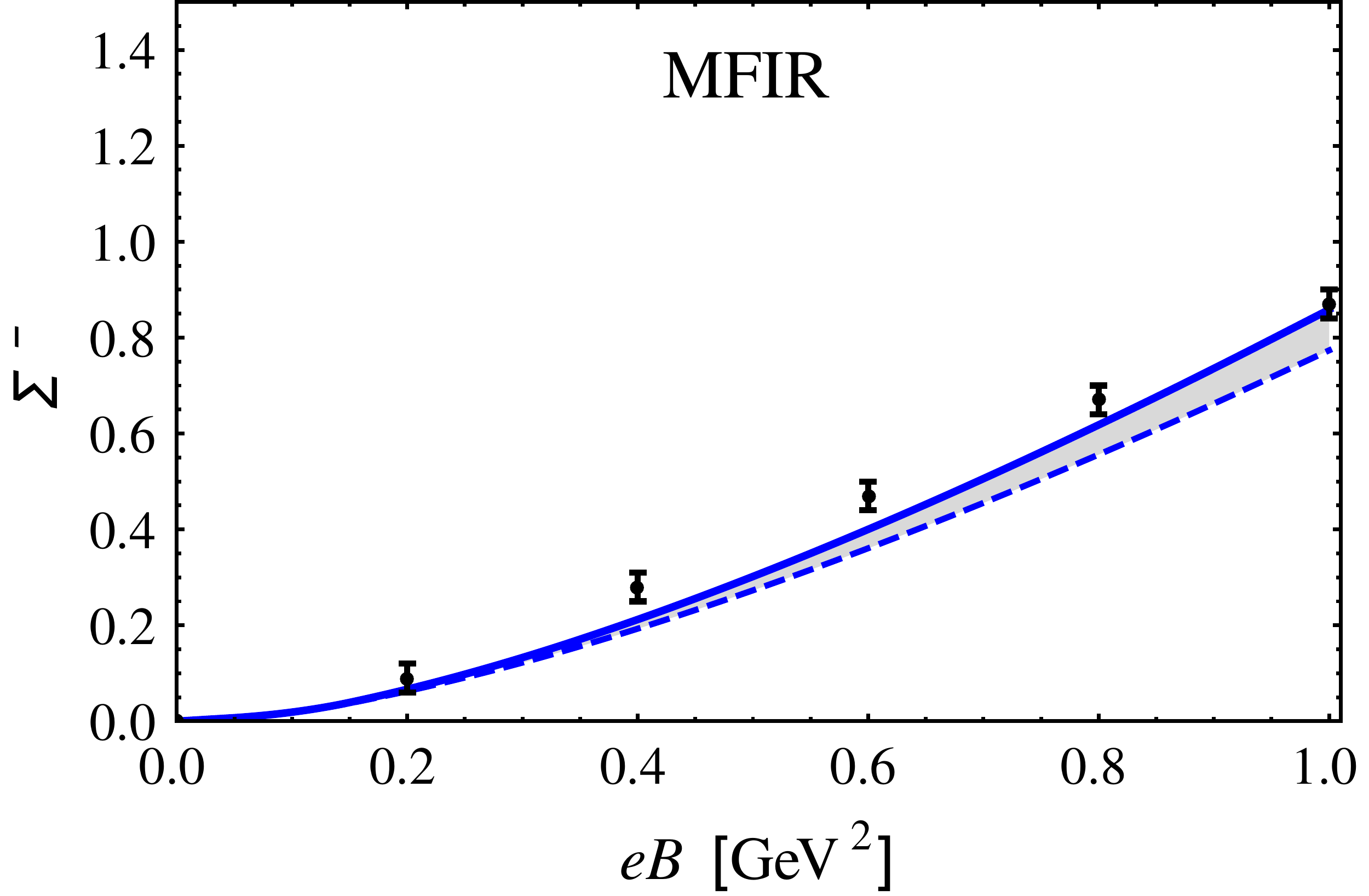}
\caption{Results for the FD form factor with $\alpha = 0.01$
compared with lattice results. Upper panels: Average flavor
condensate as a function of $eB$: nMFIR (left panel)  and
MFIR (right panel). Lower panels: difference of the up and the
down quark condensates as a function of $eB$: nMFIR (left
panel)  and MFIR (right panel)}
\label{fig4}
\end{center}
 \end{figure*}
We turn now to the form factor regularizations within the
MFIR scheme. In this case we consider the gap equation,
Eq.(\ref{gapeq3d}), where only the divergent integral $I^{3d}_1$ defined
in Eq.(\ref{I3vac}) is
regularized through the use of the several form factors just
discussed.

\begin{figure}[!t]
\includegraphics[width=0.9 \linewidth,angle=0]{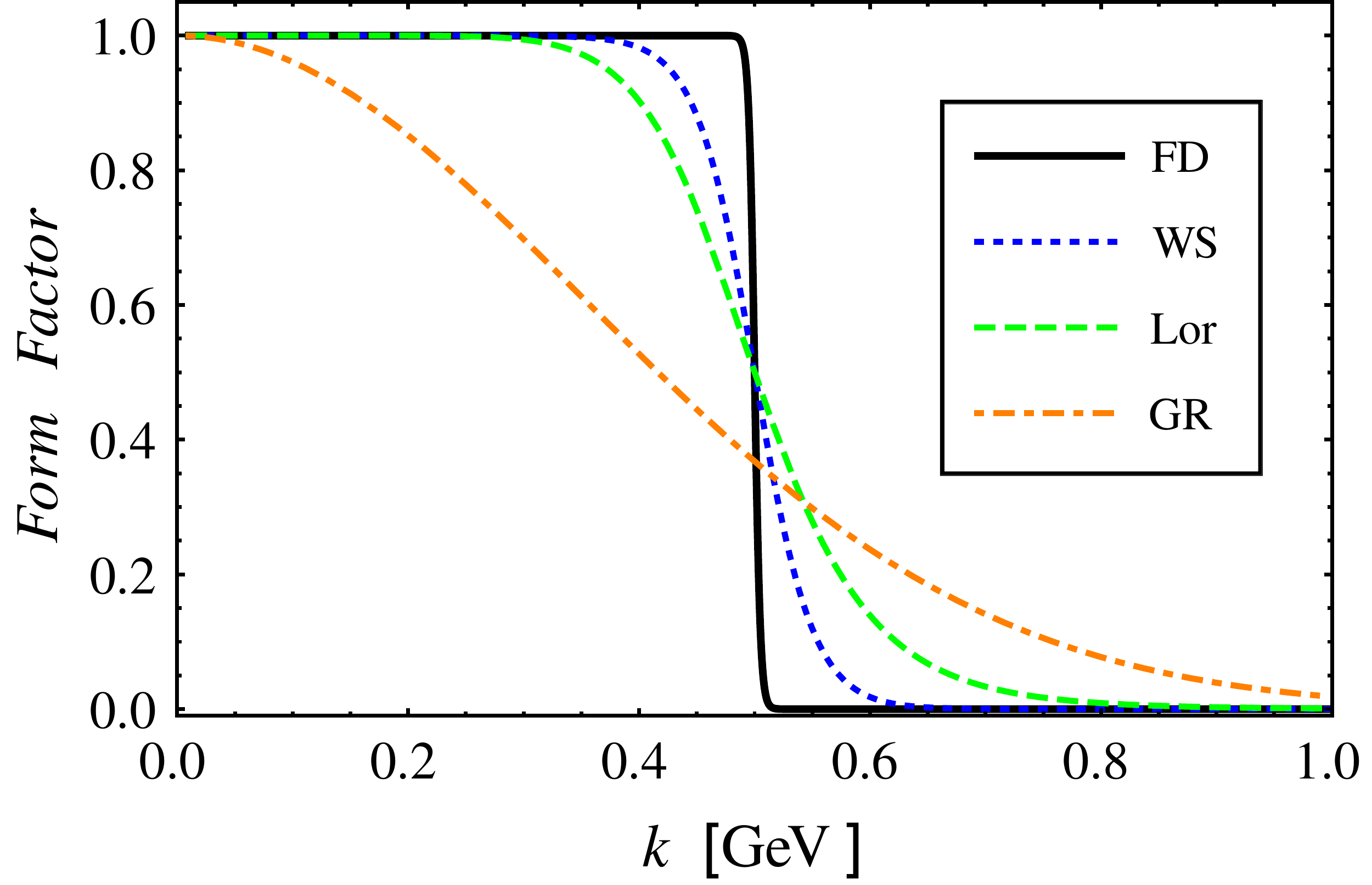}
\caption{Behavior of the form factors FD, WS, Lor5 and GR as a functions of the momentum. In this comparison we use
$\Lambda = 0.5$ GeV}.
\label{fig5}
 \end{figure}

In the Figs. \ref{fig1}-\ref{fig5}  we present our numerical
results for the behavior of the condensates as a function of $eB$
in bands for each parametrization given in Table~\ref{resfitb}.
Left panels correspond to the nMFIR scheme while those on the
right to the MFIR one. The dashed lines (solid lines) correspond
to the higher(lower) value of $-\Phi_0^{1/3}$ at $B=0$ as given in
Table~\ref{resfitb} for a particular parametrization.

Our numerical results for the average quark condensate as a function of the magnetic field
in the case of the Lor5 regulator are shown in the upper panels of Fig.~\ref{fig1} together with the lattice
results of Ref.\cite{Bali:2012zg}. To test the stability of our results we have used different parametrizations
compatible with phenomenological bounds for $\Phi_0$.

In fact, they all fall within the quite narrow band indicated in the
figure. In addition, we have performed the same calculations using the WS regulator with $\alpha=0.1$,
and we can see in Fig.2  that again the corresponding results fall basically in the same band as those
of the Lor5 regulator.
Therefore, we confirm the results reported in Fig.3 of Ref.\cite{Bali:2012zg} noting,
in addition, that they are quite insensitive to the model parametrization. One can then conclude
that the use of LorN and WS form factor regulators within the nMFIR scheme leads to a behavior of the average condensate
which is in reasonable agreement with lattice results only up to $eB \sim 0.3$ GeV$^2$.

In the Fig.~\ref{fig3} we show the condensate as a function of $eB$
in the case of the GR regulator~\footnote{For GR form factor is not possible to find a model parametrization
that satisfies the same empirical constrains that the other regularization procedures. Thus, in this case
we use $f_{\pi}=0.086$ GeV. Similar issue was previously noted in~\cite{GR_proceed}}. One interesting
aspect of using the GR form factor is that for this regulator the oscillations
that appear (in nMFIR scheme) in the behavior of the condensates using the LorN and the WS form factors are not present.
It should be noted, however, that the corresponding results for the condensates compare quite poorly with
the lattice ones.

We can see the condensate as a function of $eB$
in the case of the FD regulator from the results of Fig.\ref{fig4}. Non-physical oscillations arise with the FD form factor
and for this regulator they are stronger than the oscillations that appear in the case of the LorN and WS
form factors. We can understand these discrepancies between different form factors analyzing the behavior of
the form factors as a function of the momentum. In Fig.\ref{fig5} we can see that FD is the sharpest
function and GR in the smoothest one and the smoothness is one factor that contributes to the magnitude of
the non-physical oscillations that appear when we use form factors. This is the reason why the Gaussian
form factor GR do not present oscillations for the quark condensates, as shown in the left panel of
Fig.\ref{fig3}.

In the case of the MFIR procedure our numerical results show that for all the different shape of the form factors we obtain a good agreement with available lattice QCD calculations. 
This clearly shows the importance of implementing the separation of the purely magnetic part from the vacuum part, avoiding in this way 
the non physical oscillations that are present in nMFIR scheme.
\subsubsection{ MFIR - 3D sharp cutoff regularization}
In this regularization scheme the gap equation, $\frac{\partial {\cal F}}{\partial M} = 0$, follows
from Eq.(\ref{gapeq3d}). The only divergent  integral $I_1^{3D}$, Eq.(\ref{I3vac}), is
regularized introducing a non-covariant cutoff $\Lambda$ as shown in Eq.(\ref{i13d}) of the Appendix~\ref{app2}.
In Fig.~\ref{fig6} we can see that this regularization procedure leads to a behavior of the average
condensate which is compatible with lattice results and very similar with that ones obtained using form factors
(Fig.\ref {fig1}-Fig.\ref{fig4}).
 There the upper bound of the band corresponds to
$(\bar \Phi_0)^{1/3}=-260$~MeV while the lower to $(\bar
\Phi_0)^{1/3}=-241$~MeV. We see that this band covers the lattice
points.

\begin{figure*}[!t]
\begin{center}
\includegraphics[width=0.45 \linewidth,angle=0]{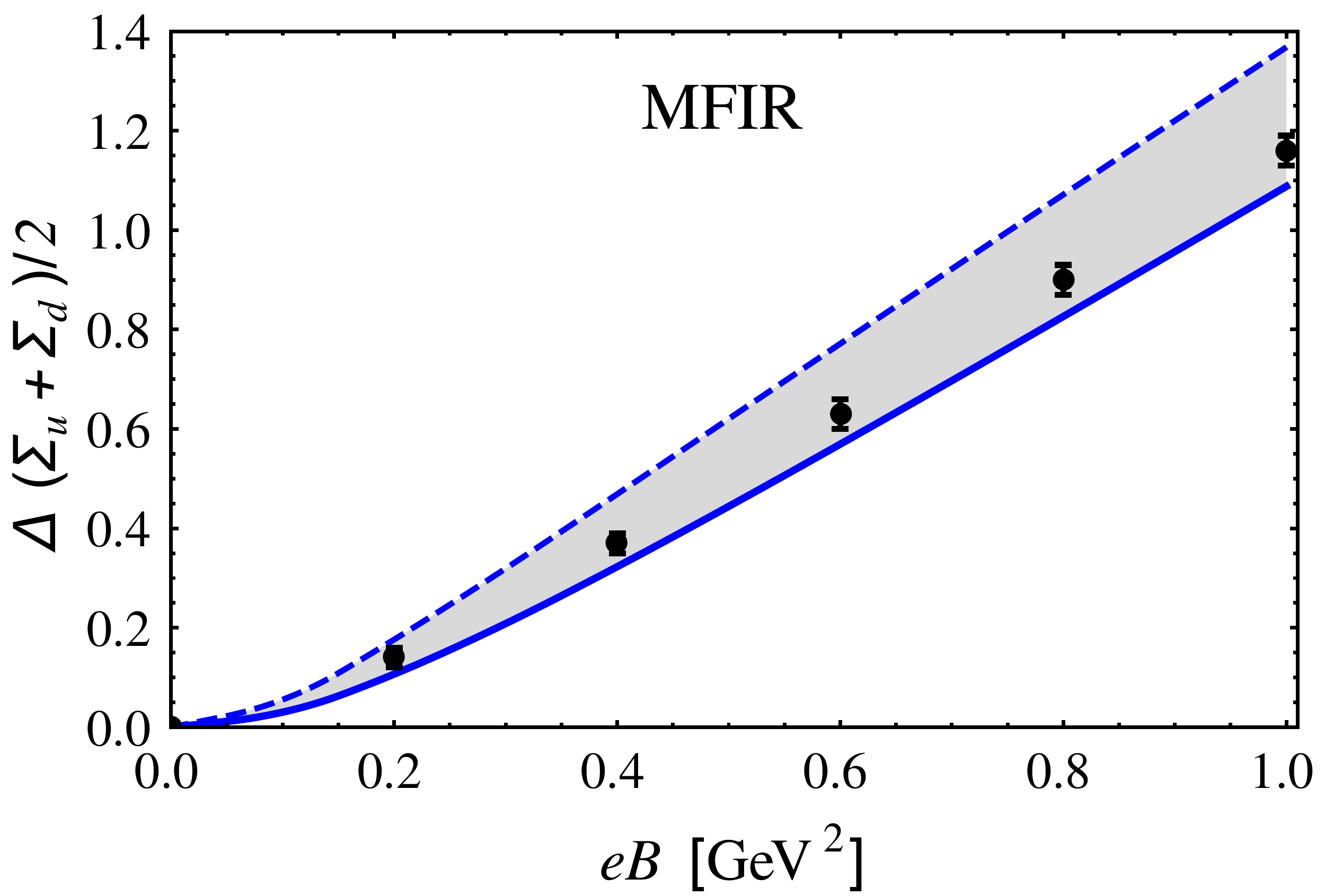}
\includegraphics[width=0.45 \linewidth,angle=0]{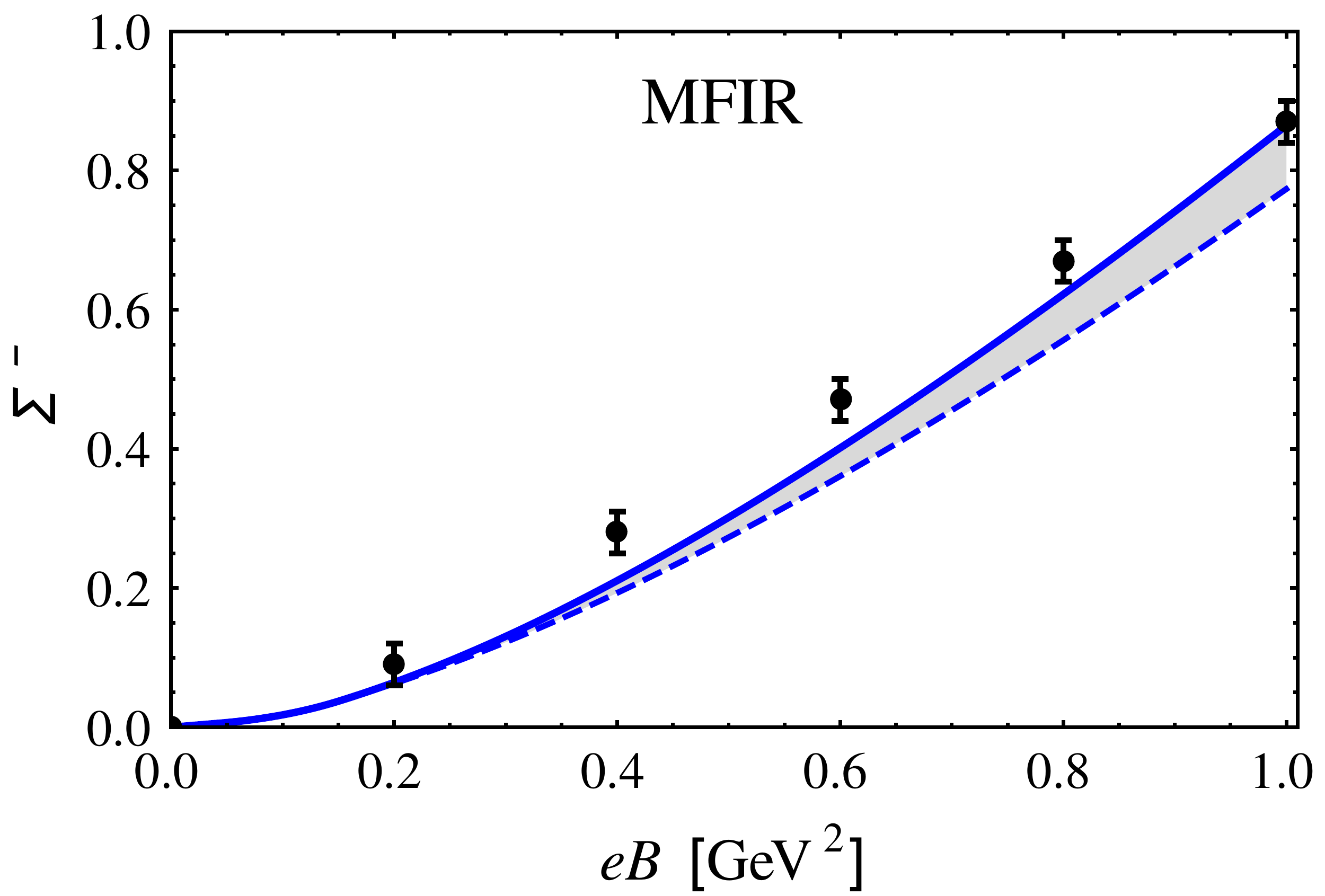}
\caption{Average of the flavor condensate as a function of
$eB$ (left panel) and difference of the up and the down quark condensates as a function of
$eB$ (right panel) evaluated with MFIR using 3D cutoff method compared with lattice results.}
\label{fig6}
\end{center}
 \end{figure*}

The difference between the condensates in this regularization method can be calculated with Eq.~\ref{dif}, using the following definition for the condensate (magnetic part)

\begin{eqnarray}
 \Phi_{B}^f=-2N_c\frac{M^3}{8\pi^2}\eta(x_f)\label{phi3d}
  \label{magpartcond}
\end{eqnarray}
where $\eta(x_f)$ is given by Eq.~\ref{eta}.

Notice that for every regularizarization based in the MFIR scheme the pure magnetic part (finite) of the condensate
is given by Eq.~\ref{magpartcond}.

\subsection{Covariant regularizations}

\begin{figure*}[!t]
\begin{center}
\includegraphics[width=0.45 \linewidth,angle=0]{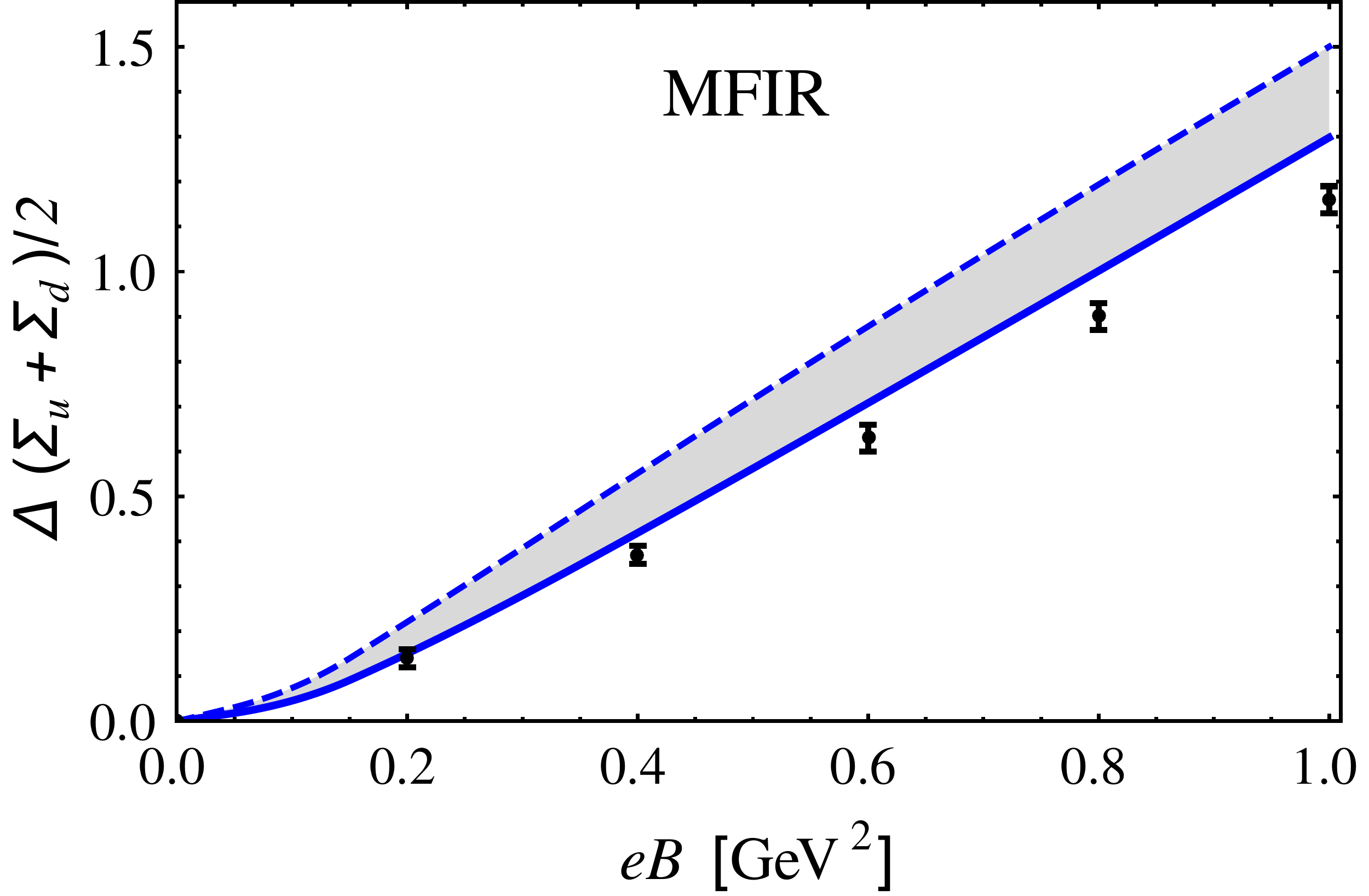}
\includegraphics[width=0.45 \linewidth,angle=0]{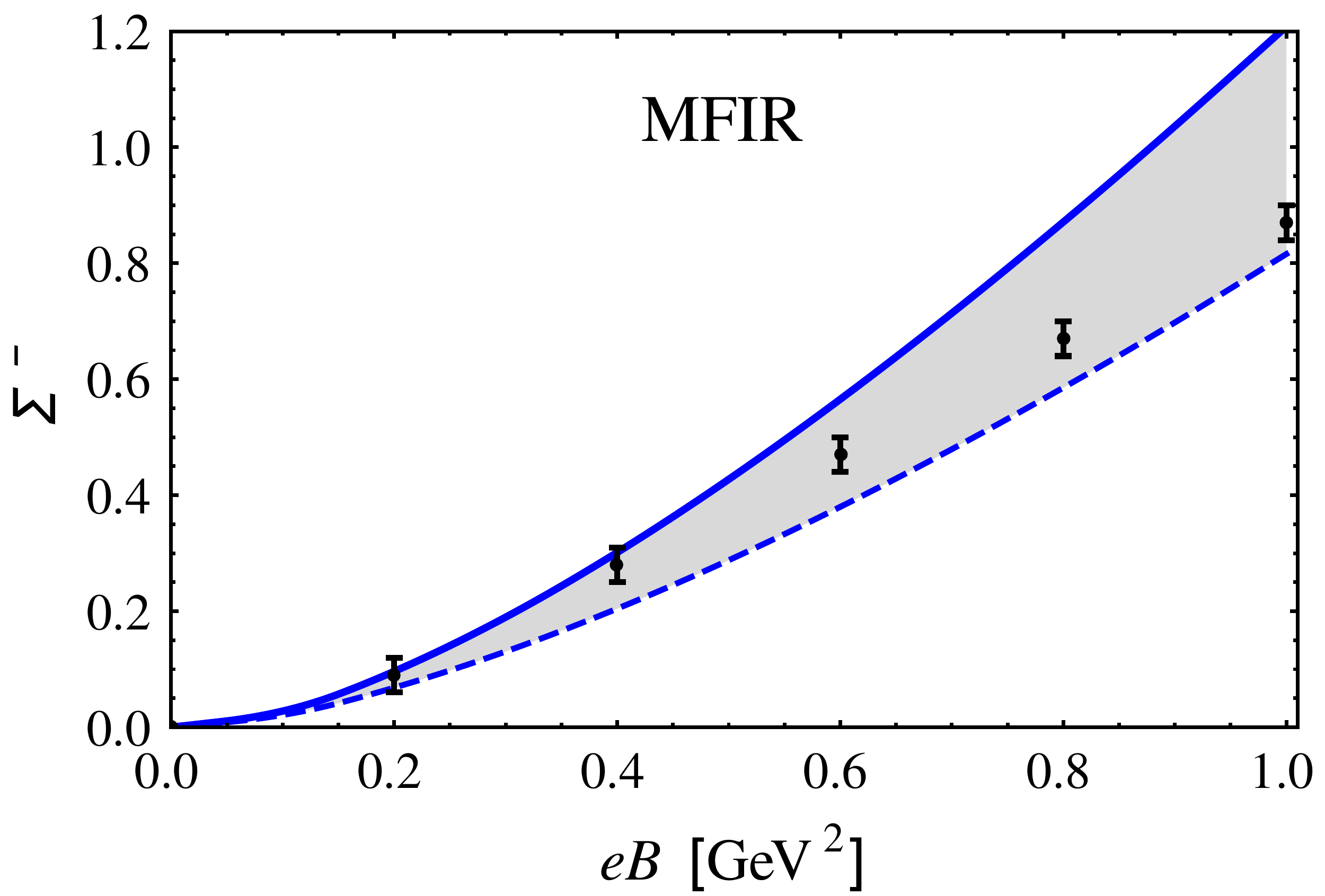}
\caption{Average of the flavor condensate as a function of
$eB$ (left panel) and difference of the up and the down quark condensates as a function of
$eB$ (right panel) evaluated with MFIR using 4D cutoff method compared with lattice results.}
\label{fig7}
\end{center}
 \end{figure*}

As examples of these covariant regularization methods we consider the 4D sharp cutoff,
proper time and Pauli-Villars. The corresponding expressions for $I_0^{4D}$, $I_0^{PT}$
and $I_0^{PV}$ as well as the associated parametrizations are given in Appendix~\ref{app2}. The
corresponding results for $\Delta(\Sigma_u+\Sigma_d)/2$ and $\Sigma^-$ using a 4D
sharp cutoff as a function of the magnetic field in comparison to those of the lattice are
shown in Fig.\ref{fig7}.

Comparing to the results in Fig.\ref{fig6} we can see that the result for the difference
$\Sigma^-$ using 4D sharp cutoff is more compatible with lattice results than the one
obtained using 3D sharp cutoff. On the other hand we note that the band associated
to the average condensate in the region $220$  MeV $< - \bar \Phi_0^{1/3} <  260$
MeV is somewhat above the lattice values.

Alternatively, proper-time was also proposed~\cite{PT} and in the nMFIR scheme the integration $I_f$
is given by:

\begin{eqnarray}
 I_f=\frac{1}{8\pi^2}\int_{\frac{1}{\Lambda^2}}^{\infty}ds\frac{e^{-sM^2}}{s}|q_f|B \coth(|q_f| Bs )
\end{eqnarray}

and the condensate by:

\begin{eqnarray}
\phi^{f}_B=-\frac{MN_c}{4\pi^2}\int_{\frac{1}{\Lambda^2}}^{\infty}ds\frac{e^{-sM^2}}{s}|q_f|B \coth(|q_f| Bs).
\end{eqnarray}

\begin{figure*}[!t]
\begin{center}
\includegraphics[width=0.45 \linewidth,angle=0]{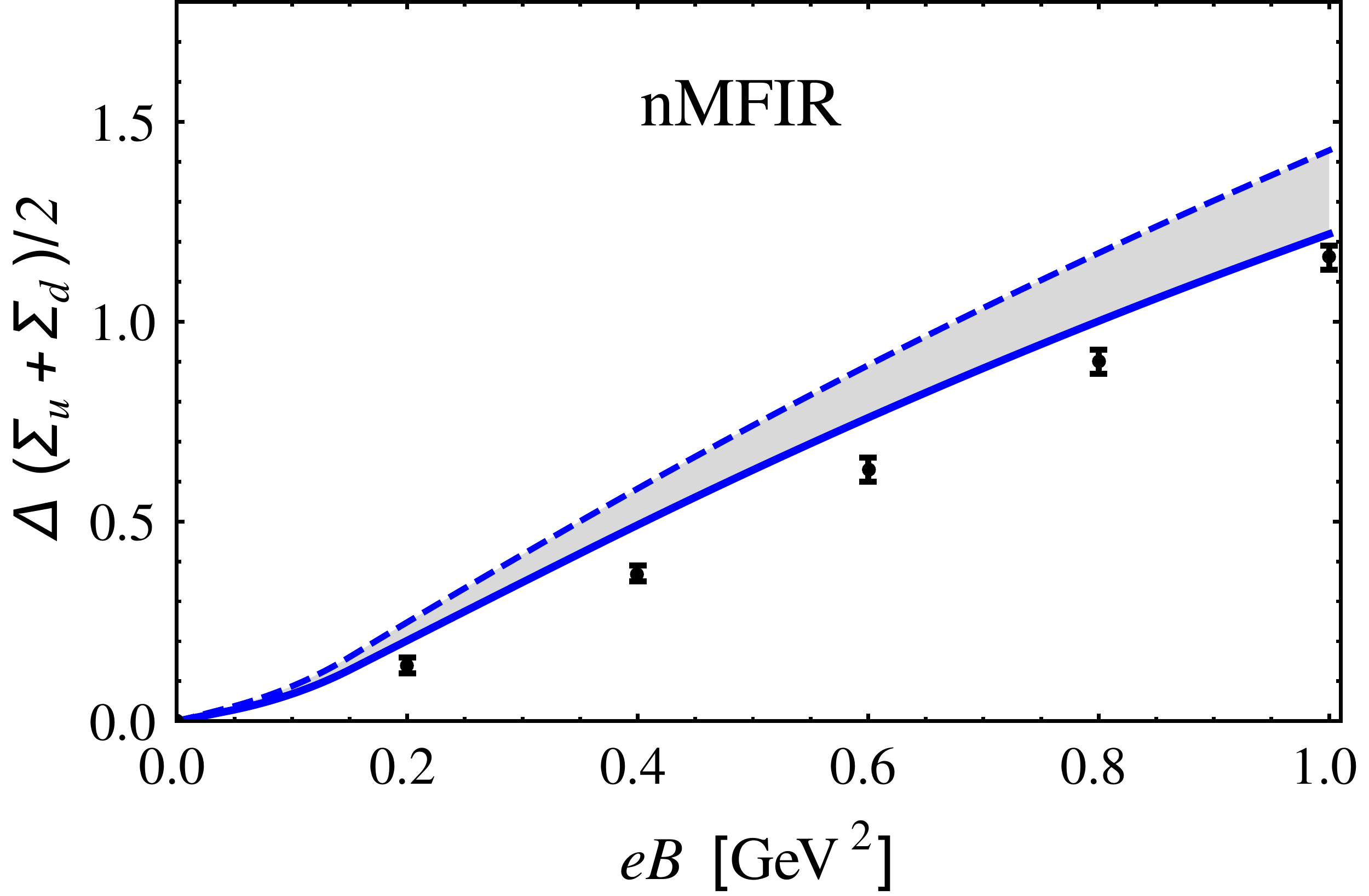}
\includegraphics[width=0.45 \linewidth,angle=0]{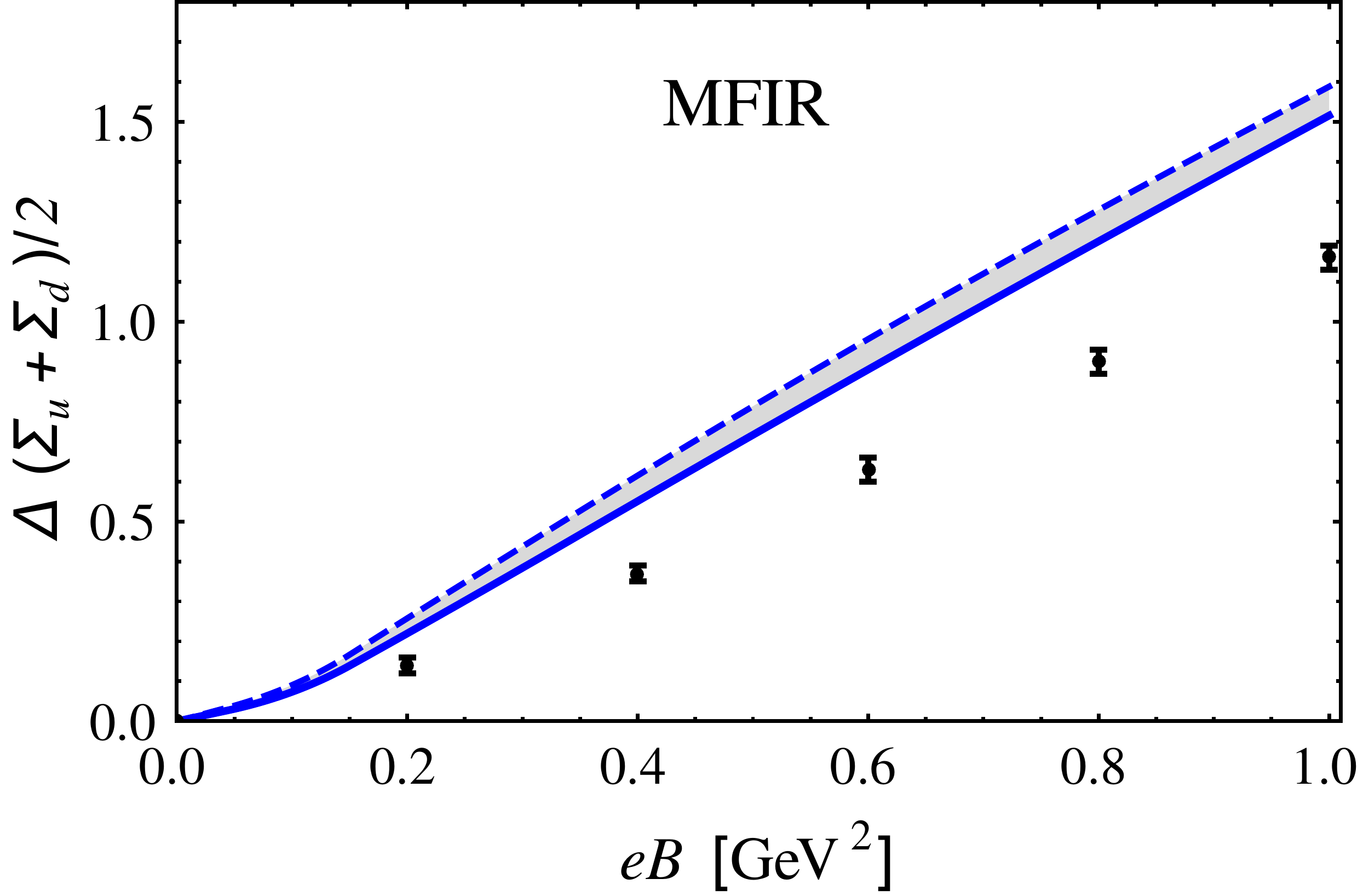}
\includegraphics[width=0.45 \linewidth,angle=0]{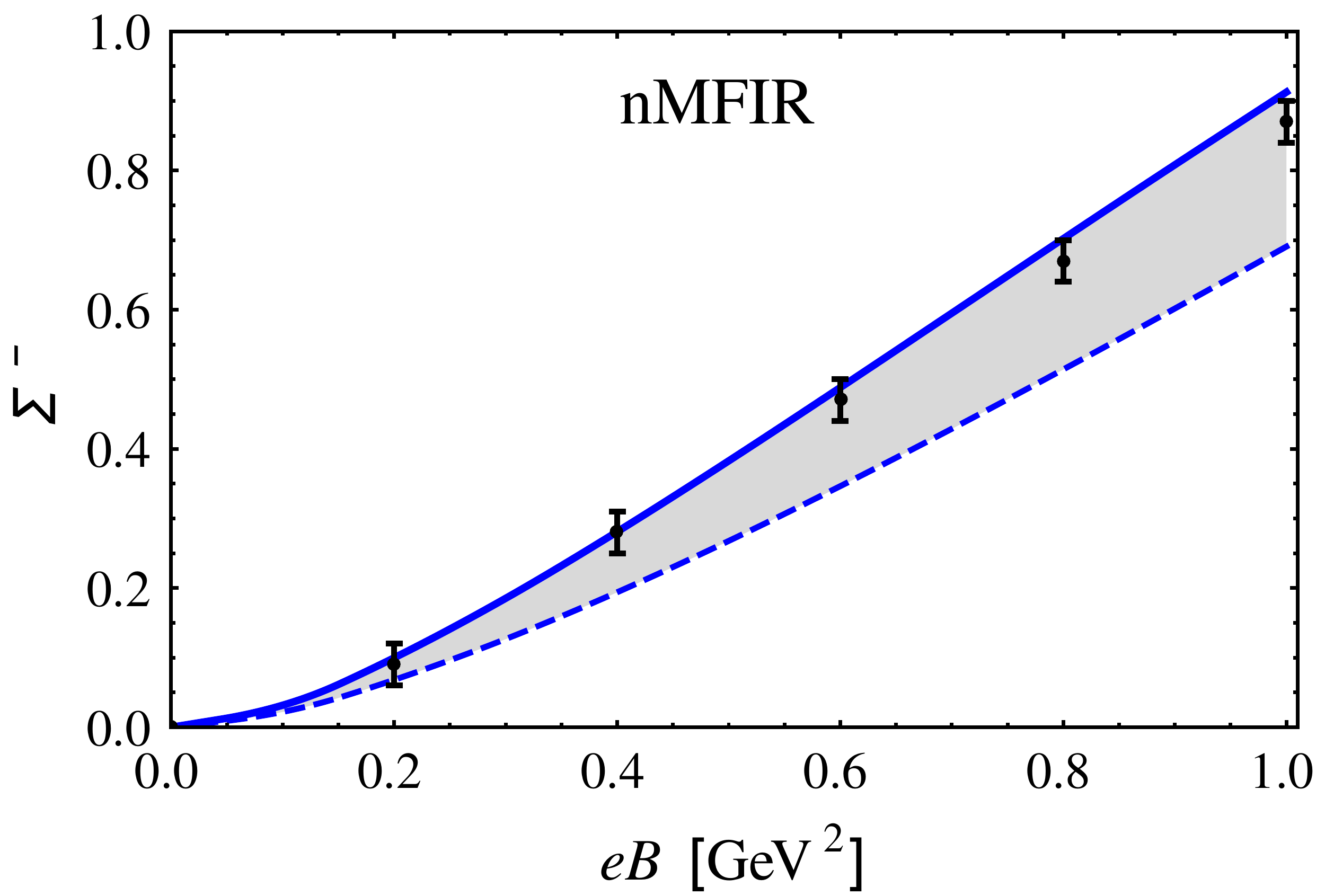}
\includegraphics[width=0.45 \linewidth,angle=0]{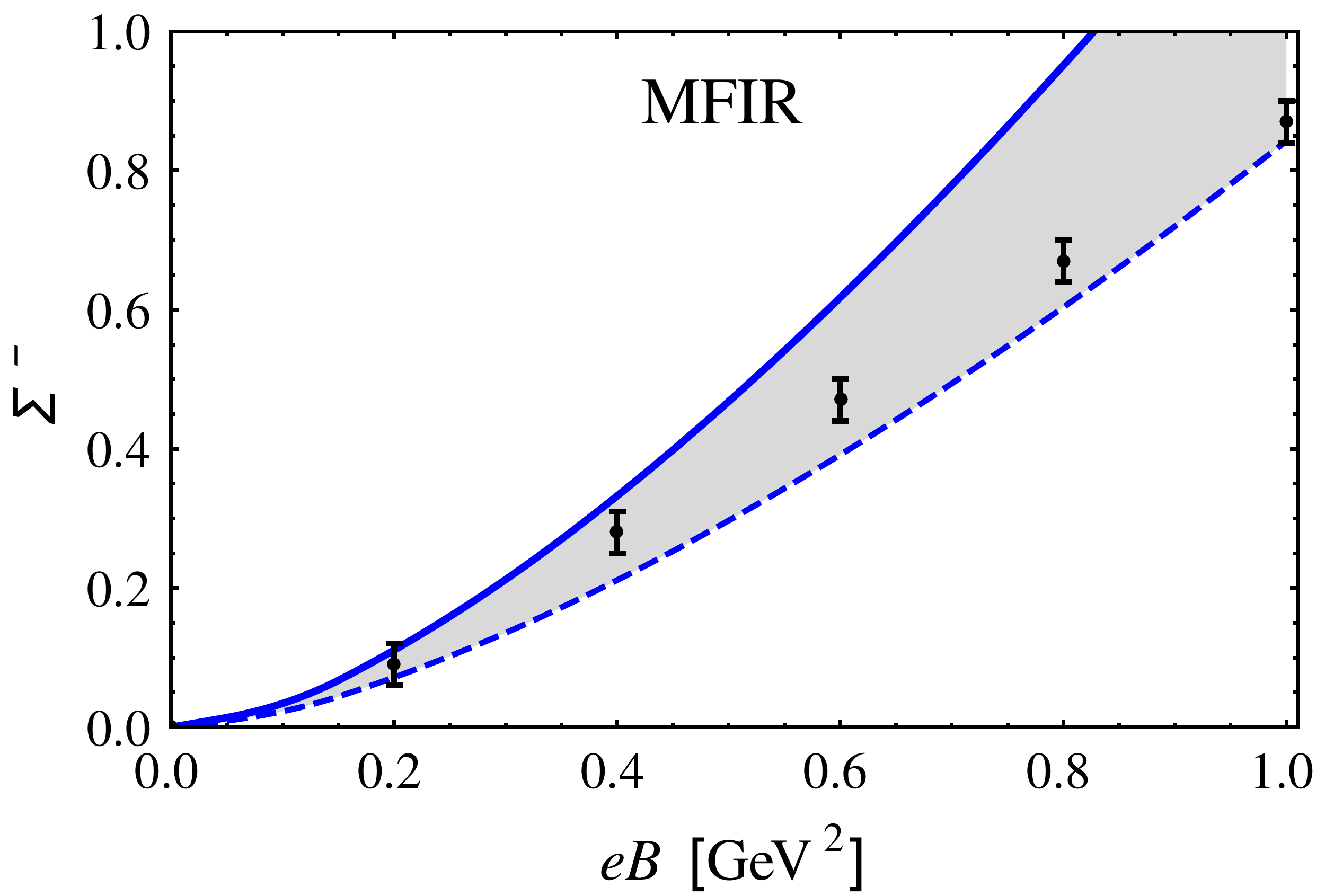}
\caption{Results for the PT method compared with lattice results. Upper panels: Average flavor condensate as a function of $eB$: nMFIR (left panel)  and
MFIR (right panel). Lower panels: difference of the up and the
down quark condensates as a function of $eB$: nMFIR (left
panel)  and MFIR (right panel).}
\label{fig8}
\end{center}
 \end{figure*}

 In this sense it is interesting to note that using the relations given in
Appendix~\ref{app1} the quantity $I_f$ to be used in the MFIR scheme can be also casted into the form
\begin{eqnarray}
I_f = \frac{1}{8 \pi^2} \int_0^\infty \frac{ds}{s^2} \ e^{- s M^2}
\Big[ |q_f| B s \ \coth{(|q_f| B s)} -1 \Big]\nonumber\\
\label{ifpt}
\end{eqnarray}
which is the magnetic term obtained within the Schwinger formalism. In Fig.~\ref{fig8} we show our
results using proper-time scheme,  we can see that the results are similar to
lattice results only for small values of $eB$.

Other very interesting regularization scheme that is used in the literature is the Pauli-Villars
regularization (PV)~\cite{revKlevansky,volkov,hatsuda,buballa,ruggieri}. In the MFIR scheme, we have to modify the integral $I_1$ in Eq.~\ref{I1vac} as
given in Eq. \ref{I1pv}. Alternatively, recent investigations focusing on
the study of the effects produced by a magnetic field in quark matter are using
PV regularization~\cite{Cao1,Cao2,Mao1,Mao2},
but they do not implement the separation of the magnetic effects from the vacuum, i. e., they use a nMFIR
procedure.

Following the prescriptions \cite{Cao1,Cao2,Mao1,Mao2}, Eq.~\ref{if} may alternatively be written
replacing the integrations as
\begin{eqnarray}
 \sum_{k=0}^\infty\alpha_k \int\frac{dp_3}{2\pi}F(E_f)\rightarrow \sum_{i=0}^2 C_i\sum_{k=0}^\infty\alpha_k \int\frac{dp_3}{2\pi}F(E_f,i),\nonumber \\
\end{eqnarray}
 also, we must introduce the regularized masses $M \rightarrow M^2_i=M^2+b_i\Lambda^2$ in the quark energy $E_{f}$. This procedure obviously rebuilds the results at $eB=0$,
but does not separate explicitly the cutoff from purely magnetic contribution.
The coefficients $C_i$ and $b_i$ are determined by the constraints $\sum_{i=0}^2 C_i=0$ and $\sum_{i=0}^2 C_iM_i^{2}=0 $ with $b_0=0$, $C_0=1$ as indicated in
\cite{revKlevansky}.

The condensate in this scheme is given by

\begin{eqnarray}
 \Phi_{B}^f&=&-2N_cM\sum_{i=0}^{2}C_i\frac{|q_f|B}{2\pi}\sum_{k=0}^{\infty}\alpha_k\nonumber\\
 &\times& \int_{-\infty}^{\infty}\frac{dp_4}{2\pi}\int_{-\infty}^{\infty}\frac{dp_3}{2\pi}
 \frac{1}{p_4^2+p_3^2+2|q_f|B+M_i^2}\nonumber \\ ,
\end{eqnarray}

\noindent and the integral $I_f$ in the nMFIR is given by

\begin{eqnarray}
I_f &=&   \sum_{i=0}^2 C_i \int_{-\infty}^\infty \frac{dp_4}{2 \pi}
\int_{-\infty}^\infty \frac{dp_3}{2 \pi} \left[\frac{|q_f|B}{2 \pi}
\sum_{k=0} \alpha_k \right.\nonumber \\
&&\times \left. \frac{1}{p_4^2+p_3^2 + 2 k |q_f| B + M^2_i}\right].
\label{ifpval}
\end{eqnarray}

\begin{figure*}[!t]
\begin{center}
\includegraphics[width=0.45 \linewidth,angle=0]{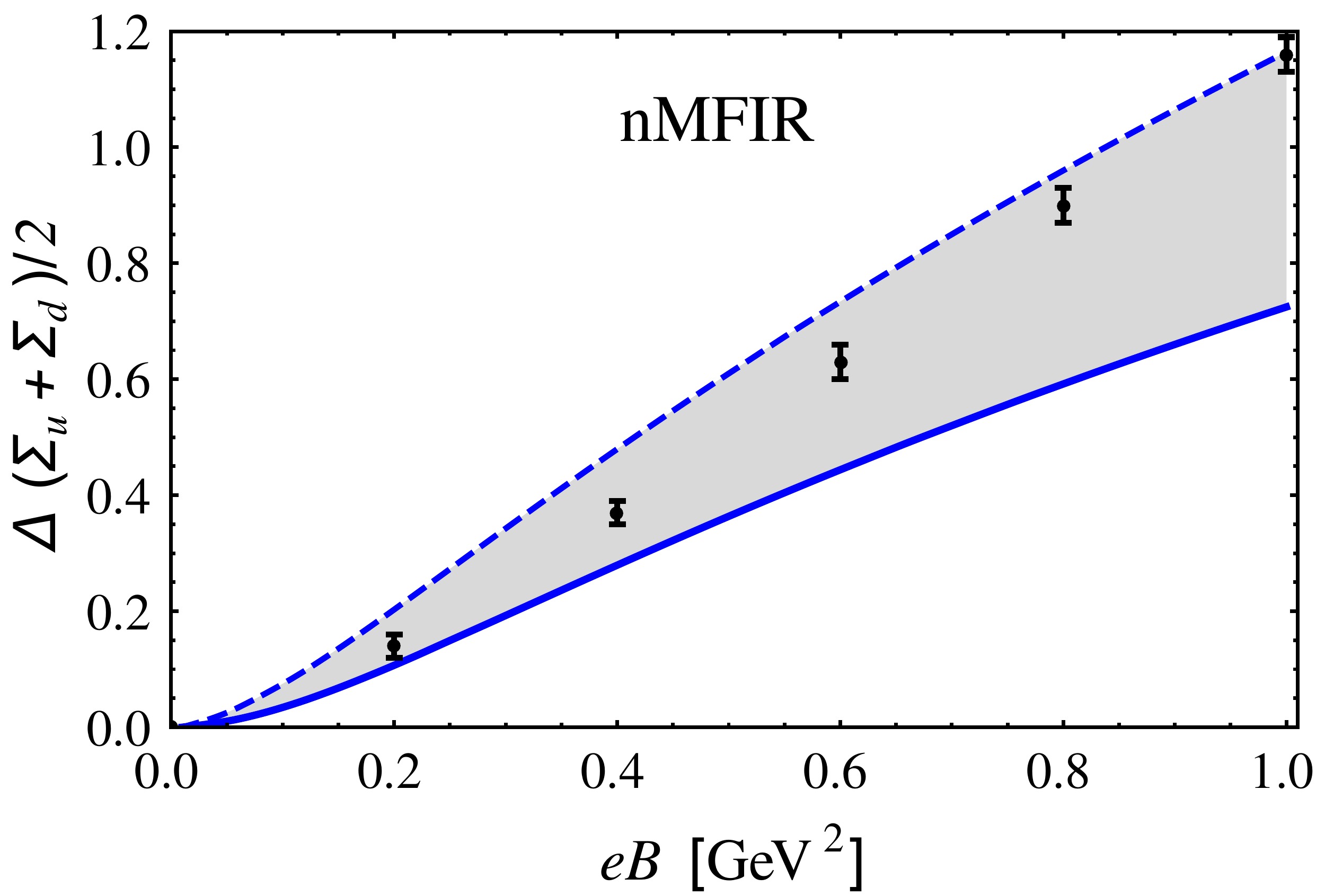}
\includegraphics[width=0.45 \linewidth,angle=0]{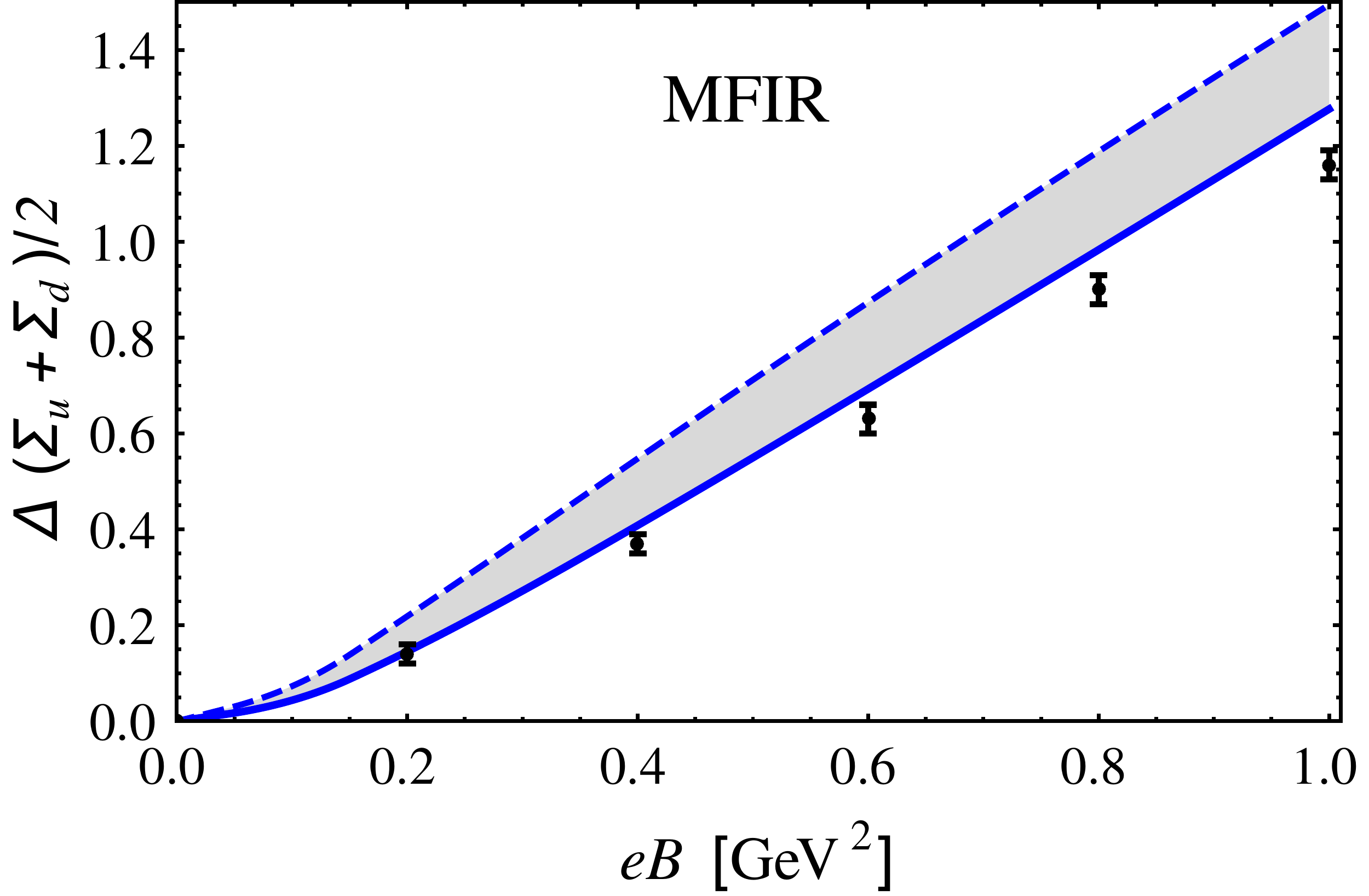}
\includegraphics[width=0.45 \linewidth,angle=0]{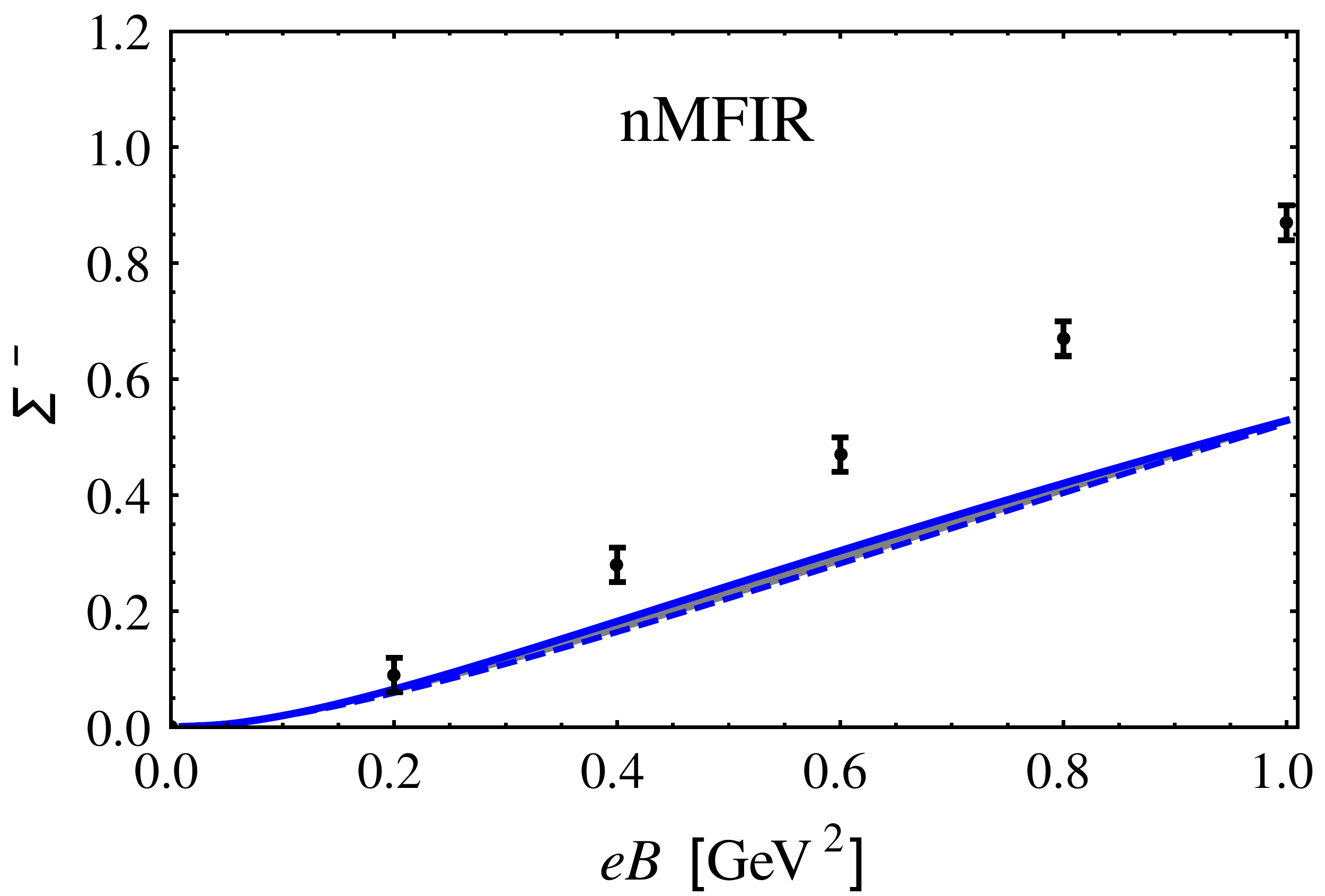}
\includegraphics[width=0.45 \linewidth,angle=0]{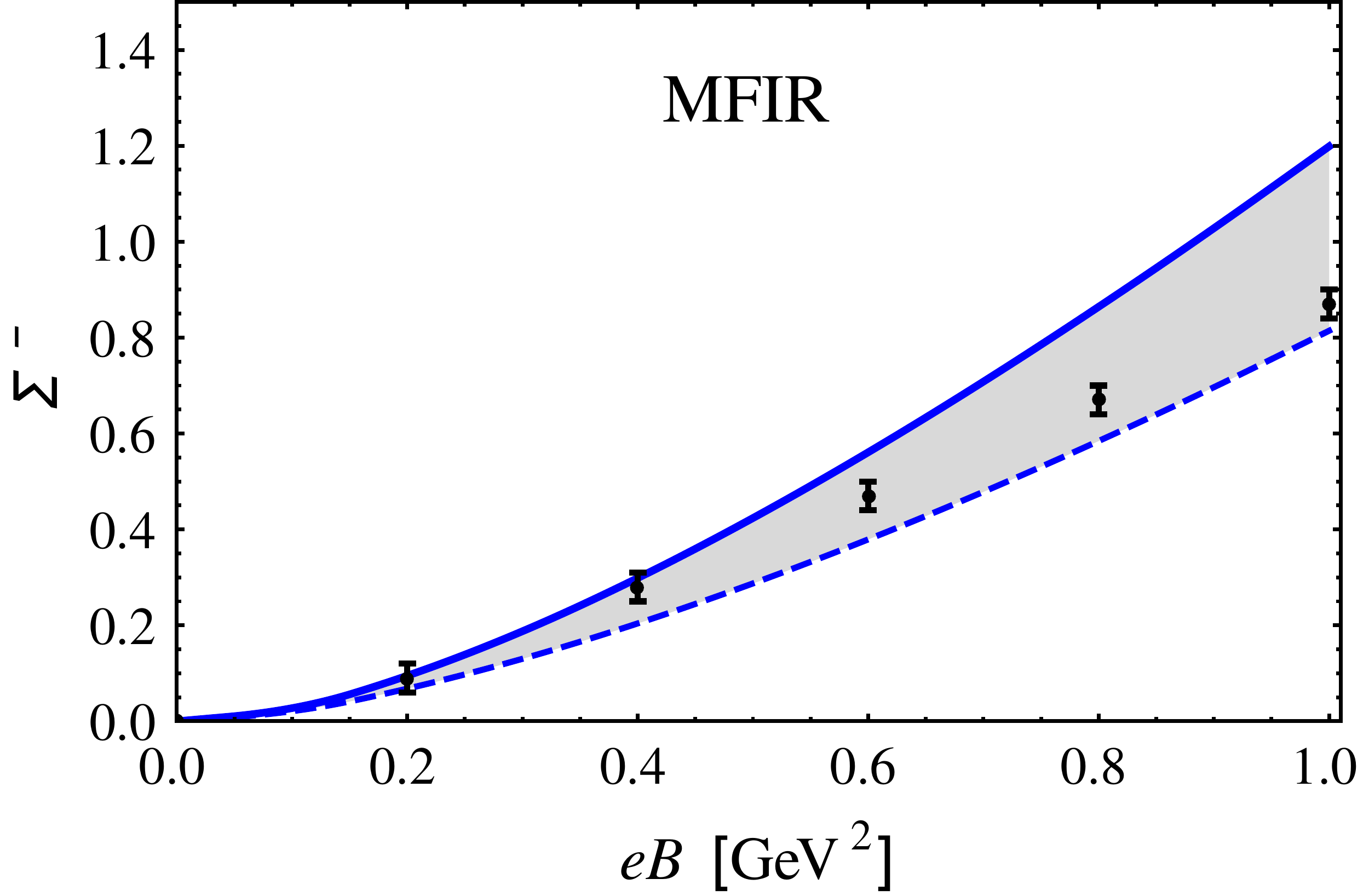}
\caption{Results for the PV method compared with lattice results. Upper panels: Average flavor condensate as a function of $eB$: nMFIR (left panel)  and
MFIR (right panel). Lower panels: difference of the up and the
down quark condensates as a function of $eB$: nMFIR (left
panel)  and MFIR (right panel).}
\label{fig9}
\end{center}
 \end{figure*}

In Fig.\ref{fig9} we compare the two procedures: PV including MFIR and without MFIR. We clearly see that if we do
not separate the magnetic contributions from the vacuum we obtain quantitative differences
compared to the case where we have used the MFIR. The results obtained using Pauli-Villars regularization with
MFIR are in agreement with lattice results.

\section{Conclusions}
\label{sec5}
In this work we use results from lattice simulations of QCD in the presence of intense magnetic fields as a benchmark
platform for comparing different regularization procedures used in the literature for the NJL type models
in the context of both the so-called ``magnetic field independent regularization'' (MFIR) scheme, where only the
non-magnetic vacuum term is regularized, and ``non magnetic field independent regularization''
(nMFIR) scheme, where also the magnetic terms are regularized.
We implement different regularization schemes in the $SU(2)$ NJL model: Form factors, Proper Time and Pauli-Villars
in both MFIR and nMFIR schemes and $3D$/$4D$ Cutoff only in the MFIR scheme.
As exhaustively discussed in this work, in the MFIR scheme for the calculation of the condensates an exact
separation of magnetic and non-magnetic vacuum contributions is performed before the adopted regularization
prescription is applied. It is important to stress that in such case only the original vacuum term of the NJL
model at $B=0$ has to be regularized.
In figures \ref{fig1}-\ref{fig4} the several non-covariant form factor regularizations
 are compared using both nMFIR and MFIR schemes. Is is seen in figures~\ref{fig1}
and \ref{fig2} that in a nMFIR scheme the Lorentzian and Woods-Saxon procedures describe approximately
the lattice data trend for the average and the difference of the flavor condensates at $eB \leq 0.3$ GeV$^2$.
 However, these figures already show the presence of a non-physical oscillatory behavior.
 The Fermi-Dirac regularization shown in Fig.~(\ref{fig4}) present a huge non-physical oscillatory behavior.
  As discussed in section \ref{sec4}, the Gaussian regulator shown in Fig.~(\ref{fig3}) has a behavior
  without oscillations but fails to satisfactorily reproduce the lattice data.  
  The form factor regularizations calculated within the MFIR scheme, shown in the right panels of
  figures (\ref{fig1}-\ref{fig4}) show a satisfactory
trend as compared to lattice results and, besides, no oscillatory behavior appears at all.
  The comparison between nMFIR with MFIR results for the form factor regularizations show
  clearly that the latter present much more consistent results when compared with lattice
  results and should be always used for
  reliable calculations of physical observables. In Fig.(\ref{fig6}) the non-covariant 3D-cutoff
  regularization is used for the calculation of the condensates in the MFIR scheme.
  It is also seen in the latter figure that a good description of the trend of the
  lattice results is achieved and no oscillatory behavior is found.

 In figures \ref{fig7}-\ref{fig9} we show results for the covariant regularizations, i. e., 4D cuttoff,
 Proper-Time and Pauli-Villars. Again we calculate the condensates using the covariant regularizations
 within the MFIR and nMFIR scheme for the Proper-Time and Pauli-Villars and only MFIR
 for the 4D-cutoff. It is obvious from these figures that the 4D-cutoff and the Pauli-Villars
  in the MFIR scheme are the best regularizations of
 all the covariant types. If one consider all the regularizations studied in this work, the conclusion
 is  that the non covariant 3D-cutoff and the covariant 4D-cutoff and Pauli-Villars  are the ones that better
 describe the lattice results for the condensates and should be chosen in any reliable calculation of
 physical quantities within the NJL model under strong magnetic fields.
  Although, in this work we focus only on the comparison of the condensates calculated within the NJL model with the
  corresponding lattice results, the MFIR scheme should be applied in the calculation
  of any physical quantity. The use of an inappropriate regularization is magnified in
  the calculation of several observables, e. g., the pion mass has been calculated in
  the literature using unreliable form factors in a nMFIR scheme and some authors have found tachyonic
  pions, huge oscillations of the pion mass which are, in fact, only an artifact
  of a bad regularization choice. Another example which highlights the importance of a correct
  regularization procedure is the calculation of thermodynamical quantities,  since
  several thermodynamic quantities involve derivatives of the thermodynamic potentials,
  they are strongly dependent on the regularization and the existence of unphysical
  oscillations would  certainly produce results completely unreliable. This is particularly 
  the case when studying the color superconducting phases in the presence of a strong magnetic 
  field where unphysical oscillations can be easily confused with actual de Haas-van Alfven oscillations.  

\section*{Acknowledgments}

This work was  partially supported by Conselho Nacional de Desenvolvimento
Cient\'{\i}fico e Tecnol\'{o}gico (CNPq) under grants 304758/2017-5 (R.L.S.F)
and 6484/2016-1 (S.S.A.),
and as a part of the project INCT-FNA (Instituto Nacional de Ci\^encia e Tecnologia -
F\'{\i}sica Nuclear e Aplica\c c\~oes) 464898/2014-5 (SSA)
and Coordena\c c\~ao de Aperfei\c coamento de Pessoal de N\'ivel Superior (CAPES) (W.R.T) 
- Brasil (CAPES)- Finance Code 001. NNS acknowledges support by CONICET and ANPCyT (Argentina), under
grants PIP17-700 and PICT17-03-0571.
%

\appendix

\section{DERIVATION OF EQ.(\ref{iffinal})}
\label{app1}
We start from the definition of $I_f$ given in Eq.(\ref{if}). After substituting in
this latter equation $\alpha_k = 2 - \delta_{0k}$ and using the Riemann-Hurwitz zeta function
\begin{equation}
 \zeta (z,x) = \sum_{n=0}^\infty\frac{1}{(x+n)^z}  ~~,
\end{equation}

it can be written as:

\begin{eqnarray}
I_f &=&  \int_{-\infty}^\infty \frac{dp_4}{2 \pi}
\int_{-\infty}^\infty \frac{dp_3}{2 \pi} \left[
\frac{1}{2 \pi}  \zeta \left(1, \frac{p_3^2 + p_4^2 + M^2}{2|q_f|B} \right) \right.\nonumber \\
&-&\left.
\frac{|q_f|B}{2\pi}\frac{1}{p_3^2 + p_4^2 + M^2}
   \right.\nonumber \\
&-& 2 \int_{-\infty}^\infty \frac{dp_1}{2 \pi} \left.\int_{-\infty}^\infty \frac{dp_2}{2 \pi}
\frac{1}{p_1^2 + p_2^2 + p_3^2 + p_4^2 + M^2}\right] \nonumber \\
\label{eqb1}
\end{eqnarray}
Next, from the integral representations of the zeta  function
\begin{eqnarray}
\int_0^\infty y^{z-1} e^{-\beta y} \coth (\alpha y) &=& \Gamma (z) \left[ 2^{z-1} \alpha^{-z}
\zeta \left( z, \frac{\beta}{2\alpha}\right)\right. \nonumber\\
&-& \left.\beta^{-z} \right] ~,
\end{eqnarray}
and
\begin{eqnarray}
 \frac{1}{A} = \int_0^\infty ds ~e^{-sA} ~,
\end{eqnarray}
then Eq.(\ref{eqb1}) can be written as:
\begin{eqnarray}
I_f & = & \int_{-\infty}^\infty \frac{dp_4}{2 \pi}
\int_{-\infty}^\infty \frac{dp_3}{2 \pi} \left[ \frac{|q_f|B}{2\pi}
\int_0^\infty dy e^{-(p_3^2+p_4^2+M^2)y} \right.\nonumber\\
&\times&\left.\coth(|q_f|By) \right. \nonumber \\
 &-& 2 \int_{-\infty}^\infty \frac{dp_1}{2 \pi} \left.\int_{-\infty}^\infty \frac{dp_2}{2 \pi}
 \int_0^\infty dy e^{-(p_1^2 + p_2^2 + p_3^2 + p_4^2 + M^2)y} \right]\nonumber \\~.
\end{eqnarray}
After performing trivial Gaussian momentum integrals, one obtains the magnetic term
in the Schwinger representation:
\begin{equation}
 I_f=\frac{1}{8\pi^2}\int_0^\infty \frac{ds}{s^2} e^{-s M^2} \left[ ~|q_f|B s \coth(|q_f|Bs) -1 \right]~.
\end{equation}
This latter integral can be calculated analytically, first we make a change of variables and write:
\begin{eqnarray}
 I_f&=&\frac{|q_f|B}{8\pi^2} \lim_{\epsilon \to 0} \left\lbrace
 \int_0^\infty ~ds e^{-s \frac{M^2}{|q_f|B}}
 \left[ ~s^{-1+\epsilon} \coth(s)\right.\right.\nonumber\\
 &-&\left.\left.s^{-2+\epsilon} \right]  \right\rbrace   ~. \label{intif}
\end{eqnarray}
Finally, using the expressions given in
the appendix of Ref.\cite{Dittrich} for the integrals involved in Eq.(\ref{intif}):
\begin{eqnarray}
\int_0^\infty ds ~e^{-s\frac{M^2}{|q_f|B}} s^{-1+\epsilon} \coth(s) &=&
 - \frac{2 x_f}{\epsilon} + 2 x_f(C+\ln 2) \nonumber\\
 &+& 2 \ln \Gamma (x_f) - \ln (2\pi) \nonumber \\
 &+& \ln x_f  ~, \\
 \int_0^\infty ds ~e^{-s\frac{M^2}{|q_f|B}} s^{-2+\epsilon} &=&
  - \frac{2 x_f}{\epsilon} +2x_f \ln (2x_f) \nonumber \\
  &+& 2x_f (C-1) ~,
\end{eqnarray}
where $C$ denotes the Euler constant, one
easily obtains:
\begin{eqnarray}
 I_f &=& \frac{M^2}{8\pi^2} \eta (x_f) \nonumber\\
 &=&  \frac{M^2}{8\pi^2} \left[
 \frac{\ln \Gamma(x_f)}{x_f} - \frac{\ln{2\pi}}{2x_f} + 1 - \left(1-\frac{1}{2x_f}\right) \ln x_f \right]
 \nonumber \\~.
\end{eqnarray}

\section{MODEL PARAMETRIZATIONS}
\label{app2}

The expression required to determined these quantities can be
written in terms of two integrals, $I_1$ and $I_2(q^2)$, whose
explicit forms are regularization dependent. At the mean field level
the gap equation leads to
\begin{equation}
M = m_0 + 4 G M N_c I_1
\end{equation}
Where $I_1$ is given by:
\begin{eqnarray}
I_1  = 4 \int \frac{d^4p}{(2\pi)^4} \frac{1}{p^2 + M^2}
\end{eqnarray}
and the average condensate is given by Eq.(\ref{cond}).
At the quadratic level the equation for the pion mass is
\begin{equation}
1 - 2 G J(- m_\pi^2) = 0  ~,
\end{equation}
where $J(q^2) = 2 N_c \left[ I_1 + q^2 I_2(q^2) \right]$, where $I_2(q^2)$ is given by
\begin{eqnarray}
I_2(q^2)=-2\int_0^1dz\int\frac{d^4p}{(2\pi)^4}\frac{1}{(p^2+M^2-z(z-1)q^2)^2}\label{I2} \nonumber ~. \\
\end{eqnarray}
Finally, the pion
decay constant is
\begin{equation}
f_\pi = -2\ Z_\pi^{1/2}\ M\ N_c \ I_2(-m_\pi^2) ~,
\end{equation}
where $Z_\pi^{-1} = - d J(q^2)/d q^2|_{q^2=-m_\pi^2}$.

Introducing the dimensionless quantities $M_\Lambda = M/\Lambda$ and $q_\Lambda = q/\Lambda$
the explicit expressions of $I_1$ and $I_2(q^2)$ are as follows.
For the case of 3D sharp cutoff we have
\begin{eqnarray}
I^{3D}_1 &=& \frac{\Lambda^2}{2 \pi^2} \left[ \sqrt{ 1 + M_\Lambda^2 }
+ M_\Lambda^2 \ln{ \frac{ M_\Lambda}{1 + \sqrt{1 + M_\Lambda^2}}}\right]\nonumber \\
\label{i13d}
\\
I^{3D}_2(q^2) &=&
\frac{1}{4 \pi^2} \int_0^1 dz
\left[ \frac{1}{\sqrt{1+ M_\Lambda^2 - z(z-1) q_\Lambda^2}} \right.\nonumber \\
&&+ \left.
\ln{ \frac{  {\sqrt{M_\Lambda^2 - z(z-1)\ q_\Lambda^2}}}{1 + \sqrt{1 +  M_\Lambda^2 - z(z-1)\ q_\Lambda^2}}}
\right]\ ,
\end{eqnarray}
while the 3D form factor regularization read
\begin{equation}
I^{FF}_1 = \frac{\Lambda^2}{\pi^2} \int_0^\infty du \frac{u^2 U_\Lambda(u^2) }{\sqrt{u^2 + M_\Lambda^2}}
\end{equation}
\begin{equation}
I^{FF}_2(q^2) =
-\frac{1}{ 4\pi^2}
\int_0^\infty du
\frac{u^2 U_\Lambda(u^2) }{ (u^2 + M_\Lambda^2 -z(z-1)q_\Lambda^2)^{3/2}} ~.
\end{equation}
For proper time regularization one gets
\begin{eqnarray}
I^{PT}_1 &=& \frac{\Lambda^2}{4 \pi^2} E_2\left(M_\Lambda^2\right)
\\
I^{PT}_2(q^2) &=& - \frac{1}{8 \pi^2} \int_0^1 dz \
E_1\left(M_\Lambda^2 - z(z-1)\ q_\Lambda^2\right)\nonumber ~, \\
\end{eqnarray}
where $E_n(x) = \int_1^\infty dt\ t^{-n} \exp{(-t x)}$ is the
exponential integral function.

For $4D$ cutoff regularization one gets
\begin{eqnarray}
I^{4D}_1 &=& \frac{\Lambda^2}{4 \pi^2}
\left[ 1 + M_\Lambda^2 \ln \frac{ M_\Lambda^2}{1+ M^2_\Lambda} \right] ~,
\\
I^{4D}_2(q^2) &=&  \frac{1}{8 \pi^2} \int_0^1 dz
\left[ \frac{1}{1+ M_\Lambda^2 - z(z-1) q_\Lambda^2} \right.\nonumber \\
&+& \left.\ln{ \left(\frac{  M_\Lambda^2 - z(z-1)\ q_\Lambda^2}
{1 +  M_\Lambda^2 - z(z-1)\ q_\Lambda^2}\right)}
\right] ~.
\end{eqnarray}

Finally, for Pauli-Villars regularization one gets
\begin{eqnarray}
I^{PV}_1 &=&\frac{\Lambda^2}{4\pi^2}\left[\left(2+M^2_{\Lambda} \right )\log{\left(1+2M^{-2}_{\Lambda} \right)} - 2\left(1+M^2_{\Lambda} \right )\right.\nonumber \\
&\times& \left.\log{\left(1+M^{-2}_{\Lambda}\right)} \right]\label{I1pv} \\
I^{PV}_2(q^2) &=&  -\frac{1}{8\pi^2}\int_0^1 dz\left[2\log \left(1+\frac{1}{M_{\Lambda}^2-z(z-1)q_{\Lambda}^2} \right) \right.\nonumber \\
&-& \left.
 \log \left(1+\frac{2}{M_{\Lambda}^2-z(z-1)q_{\Lambda}^2} \right) \right].
\end{eqnarray}
where, here, $M_\Lambda = M/\Lambda$. 

\end{document}